\documentclass[a4paper,11pt]{article}
\pdfoutput=1

\usepackage{jheppub}
\usepackage[T1]{fontenc}
\usepackage[utf8]{inputenc}
\DeclareUnicodeCharacter{FFFC}{}

\usepackage{mathrsfs}
\usepackage{amsfonts}
\usepackage{epsfig}
\usepackage{amsthm}
\usepackage{amsmath}
\usepackage{amssymb}
\usepackage{graphicx}
\usepackage{verbatim}
\usepackage{booktabs}
\usepackage{subcaption}
\usepackage{array,multirow}
\usepackage{url}
\usepackage{color}
\usepackage{cprotect}
\usepackage{tabularx}
\usepackage{dcolumn}
\usepackage{bm}
\usepackage{slashed}
\usepackage{physics}
\usepackage{framed}
\usepackage{orcidlink}
\usepackage{enumitem}
\usepackage{tikz}
\usetikzlibrary{arrows.meta,positioning,fit}

\definecolor{nicered}{rgb}{0.5,0.,0.}
\definecolor{nicegreen}{rgb}{0.,0.5,0.}
\definecolor{niceblue}{rgb}{0.,0.,0.5}
\hypersetup{colorlinks,citecolor=nicegreen,linkcolor=nicered}

\setlist{nolistsep}

\newcommand{\AdS}{\mathrm{AdS}}
\newcommand{\CFT}{\mathrm{CFT}}
\newcommand{\BY}{\mathrm{BY}}
\newcommand{\ren}{\mathrm{ren}}
\newcommand{\eff}{\mathrm{eff}}

\newcommand{\sphere}{\mathbb{S}}
\newcommand{\Neff}{\mathcal N_{\eff}}
\newcommand{\NC}{\mathcal N_C}
\newcommand{\Sth}{S_{\mathrm{th}}}

\title{Large-$N$ Carrollian Thermodynamics from AdS Black-Hole Phase-Space Contractions}

\author[a,b,1]{Yingnan Xu\,\orcidlink{0009-0005-8296-0850}\note{Corresponding author.}}
\author[c]{Shuangshuang Chu\,\orcidlink{0009-0004-4396-582X}}

\affiliation[a]{Zhongtai Securities Institute for Financial Studies, Shandong University,\\
Jinan, Shandong 250014, China}
\affiliation[b]{Department of Physics, Southern Methodist University,\\
Dallas, Texas 75206, U.S.A.}
\affiliation[c]{School of Statistics, Dongbei University of Finance and Economics,\\
Dalian, Liaoning 116025, China}

\emailAdd{yingnanx@mail.smu.edu}
\emailAdd{shuangshchu@gmail.com}

\abstract{We develop the boundary large-rank interpretation of finite Carrollian black-hole thermodynamics and its sourced stress, scalar spectral and interacting thermal observables. Correlated scaling of the time generator and Newton coupling preserves the extended Hamiltonian differential while the generator temperature vanishes and the holographic normalization grows. The Brown--York source variation fixes the relative weights of energy density, momentum density and spatial stress, giving Ward equations on the Carrollian and thermal time scales. Boundary volume and normalization variations reproduce the bulk pressure work and characterize the Hawking--Page crossing through a chemical potential. Source coupling fixes the scalar operator map and spectral clock Jacobian. Numerical solutions in four-dimensional Schwarzschild-AdS determine angular transmission and finite-frequency absorption. In the type IIB family, the logarithmic scrambling interval imposes an additional hierarchy between rank and 't Hooft coupling. We derive a correlated limit in which the string length vanishes while a finite transverse Regge kernel survives in the normalized four-point functional. Its integrated attenuation and spatial width obey a common relation. This interacting observable distinguishes large-rank sequences with the same leading thermodynamics and asymptotic Regge growth rate. Thermal Fourier--Mellin moments retain the scalar spectrum in a contracting frequency window. Time translations shift their weights with angular source terms retained. The BTZ dictionary supplies the corresponding large-central-charge equilibrium example.}


\begin{document}
\maketitle

\section{Introduction}
\label{sec:intro}

Finite Carrollian black-hole energy and a vanishing generator temperature require a joint specification of the boundary clock and the number of degrees of freedom. Their correlated scaling determines which equilibrium and real-time observables remain finite. We investigate this question for neutral AdS black holes, connecting the extended Hamiltonian charge to boundary stress responses, scalar absorption and an interacting thermal four-point function. The observation interval is part of this comparison: thermal response and scrambling probe different scales of the selected clock. In this paper we determine the corresponding source normalizations and coupling hierarchies, and identify state information that survives beyond the homogeneous energy and entropy.

We use the extended Hamiltonian contraction of Ref.~\cite{Xu:2026vpj} as the thermodynamic input. Its static boundary construction includes the work associated with changes of the cosmological constant and asymptotic clock sources. The charge and its source dependence determine the normalization used in the large-rank boundary description. The original black-hole laws, Euclidean ensemble and Noether-charge framework supply the underlying equilibrium structure \cite{Bekenstein:1973ur,Bardeen:1973gs,Hawking:1975vcx,Gibbons:1976ue,Gibbons:1977mu,Wald:1993nt,Iyer:1994ys,Iyer:1995kg}. We identify its boundary realization through the holographic correspondence \cite{Maldacena:1997re,Gubser:1998bc,Witten:1998qj}, the renormalized gravitational stress tensor \cite{Henningson:1998gx,Balasubramanian:1999re,deBoer:1999tgo,deHaro:2000vlm,Skenderis:2002wp,Papadimitriou:2004ap}, and the Brown--Henneaux central-charge normalization \cite{Brown:1986nw}. These ingredients fix the energy unit against which the Carrollian and large-rank limits can be compared.

A change of the AdS radius has several boundary consequences. Extended black-hole thermodynamics identifies the associated bulk work through the cosmological pressure and thermodynamic volume \cite{Kastor:2009wy,Cvetic:2010jb,Dolan:2011xt,Kubiznak:2012wp,Kubiznak:2016qmn}. Karch and Robinson related the generalized Smarr relation to the overall color dependence of holographic thermodynamic quantities \cite{Karch:2015rpa}. Ahmed and collaborators formulated the boundary differential by treating the conformal factor as a thermodynamic variable \cite{Ahmed:2023snm}. Related analyses developed central-charge ensembles, higher-dimensional embeddings and holographic thermodynamic potentials \cite{Johnson:2014yja,Xiao:2023lap,Frassino:2022zaz,Mann:2024sru}. These analyses identify how the number of degrees of freedom enters the thermodynamic potentials. Our calculation specifies the combination selected by the Carrollian limit: the boundary spatial volume and stress-tensor normalization vary together when the AdS scale varies at fixed contracted Newton coupling. The resulting chemical potential retains a finite value while its parent normalization grows. Generalized Euler relations \cite{Mancilla:2024spp} and quasilocal boundary pressure constructions \cite{Borsboom:2026ash} provide complementary comparisons, with their respective source prescriptions retained.

The Hawking--Page transition supplies an equilibrium reference for the response calculation \cite{Hawking:1982dh,Witten:1998zw}. Its geometric crossing is inherited from the parent ensemble, while the finite boundary free energy and the holographic chemical potential identify the associated thermodynamic variation. The numerical spectral results then distinguish states with different horizon radii through their angular absorption. In this way the equilibrium charge, local source response and thermal probe data are presented within one set of clock and coupling conventions.

Carroll geometry originates from the contraction of relativistic spacetime symmetry \cite{Levy-Leblond:1965dsc,SenGupta:1966qer}. Its clock, spatial metric and preferred temporal vector provide the source data for a degenerate boundary geometry \cite{Duval:2014uoa,Duval:2014uva,Hartong:2015xda,Bergshoeff:2022eog,Figueroa-OFarrill:2022nui,Hartong:2022lsy,Hansen:2021fxi,Campoleoni:2022ebj}. At null hypersurfaces and horizons, these data organize intrinsic evolution and surface charges \cite{Ciambelli:2019lap,Herfray:2021qmp,Ciambelli:2025unn,Donnay:2019jiz,Freidel:2022vjq,Freidel:2022bai,Redondo-Yuste:2022czg,Husnugil:2025edm,Bagchi2026}. P\'erez constructed magnetic AdS Carroll charges and their asymptotic symmetry algebra \cite{Perez:2022carroll}. Ecker and collaborators developed intrinsic Carroll black holes, thermal properties and extremal surfaces, including Schwarzschild, charged and BTZ examples \cite{Ecker:2023carroll}. These constructions supply the geometric and charge background for our boundary analysis. The present response functions are obtained from the relativistic renormalized functional along a specified clock and coupling curve, so their source weights can be followed before the limit is taken.

The microscopic interpretation depends on the limiting source prescription. Fontanella and Payne constructed a Carroll limit of the string and gauge-theory descriptions of the AdS correspondence \cite{Fontanella:2025tbs}. We consider a sequence of relativistic theories whose thermal clocks and stress-tensor normalizations are correlated. For equilibrium observables, the type IIB embedding fixes the rank dependence and suppresses string loops at fixed large 't Hooft coupling. It also specifies the meaning of the low temperature: the temperature measured by the selected generator tends to zero, while the finite-clock thermal scale is held fixed. Work on tantum gravity, quadratic-gravity contractions and nonrelativistic holography illustrates other correlated geometric and coupling limits \cite{Ecker:2024czh,Tadros:2023teq,Fontanella:2024kyl,Fontanella:2026gaq}. Each construction is characterized by the observables and source variations that remain finite.

The same type IIB sequence supports an interacting observable on a longer observation interval. Recent work constructs Carrollian correlators from doubly scaled vacuum amplitudes in the large-radius limit of the super Yang--Mills correspondence \cite{Bagchi:2026doublescaled}. We retain the AdS radius and finite temperature and examine a normalized thermal four-point function in the type IIB family. The curved-horizon Regge amplitude \cite{Shenker:2014cwa} supplies the string-scattering input. Combining its accumulated boost with our rank normalization gives a hierarchy for the Einstein scrambling limit and a marginal family with finite transverse string spreading. The resulting interaction kernel distinguishes sequences whose leading thermodynamics and limiting growth rate coincide. Its integrated attenuation is tied to its spatial width. Recent links between pole-skipping trajectories and stringy horizons \cite{Gao:2025stringy} provide a complementary perspective on the Regge information retained by this observable.

The sourced stress-tensor calculation resolves how local conservation behaves under the same contraction. We derive the energy, momentum and spatial-stress responses by varying the lapse, scaled shift and spatial metric. The local conservation equations retain a factor of the contraction parameter in their spatial terms. This factor distinguishes evolution at fixed Carrollian time from evolution on the thermal time scale. The global energy remains the constant mode of the angularly smeared Brown--York energy. Spatially dependent time shifts also transform the clock and shift sources, and their Ward identities contain those variations. In equilibrium, charge conservation governs time evolution, while the extended Hamiltonian identity governs variations between black-hole states. Both statements refer to the same global charge.

We study the thermal response of a minimally coupled scalar in the Schwarzschild-AdS geometry. Horizon flux determines its low-frequency spectral coefficient, including the transmission of a boundary angular harmonic to the horizon. Numerical solutions in four bulk dimensions quantify that transmission beyond the homogeneous channel and resolve its frequency dependence through the thermal range. The calculation supplies the normalization required to compare spectral densities in the finite Lorentzian and Carrollian clocks. The Fourier Jacobian and the operator normalization enter independently. Keeping them together produces a finite thermal Fourier--Mellin moment in a frequency window that contracts with the temperature. The incomplete Bose moment retains the dependence on the physical size of this window. We also determine the leading spatial filtering of angular modes as the black-hole radius increases.

Long and Xiao construct thermal correlators at null infinity using the real-time formalism \cite{Long:2025thermal}; Long, Qu and Xiao relate black-hole Carrollian correlators to reflection and transmission amplitudes in asymptotically flat geometries \cite{Long:2026blackhole}. Our probe has an AdS boundary source and a future-horizon absorption condition, and its thermal response belongs to a large-rank equilibrium family. This choice determines the spectral kernel and the clock dependence of the Mellin data. Thermal one-point coefficients and local operator-product coefficients enter distinct places in thermal conformal analysis \cite{Iliesiu:2018thermal}; we retain that distinction when describing the three-point transform.

Section~\ref{sec:bulk} fixes the finite branch and its large-rank normalization, including the type IIB and BTZ dictionaries. Section~\ref{sec:stress} derives the boundary sources and Brown--York charge. Section~\ref{sec:thermodynamics} resolves the extended work into boundary volume and normalization variations and determines the thermal ensemble; the associated susceptibilities are developed in Appendix~\ref{app:conventions}. Section~\ref{sec:ward} obtains the sourced Ward identities and their time scales. Section~\ref{sec:scalar} calculates the scalar spectral response, and Section~\ref{sec:regge} derives the interacting type IIB limit on the logarithmic scrambling interval. Section~\ref{sec:celestial} expresses the thermal data and time-shift identities in the Fourier--Mellin basis. Section~\ref{sec:conclusion} discusses the resulting boundary interpretation.

\section{The finite branch and large-rank normalization}
\label{sec:bulk}
\label{sec:largeN}

The boundary observables depend on a common normalization of the time generator and gravitational coupling. We establish that normalization on the neutral equilibrium family and then identify its microscopic rank or central-charge meaning. Thermodynamic variations within a parent theory and motion along the contraction sequence are kept distinct throughout.

\subsection{Hamiltonian normalization and finite thermodynamic coordinates}

We use the extended static Hamiltonian identity of Ref.~\cite{Xu:2026vpj}, with the reference energy assigned to AdS at the same cosmological constant and boundary clock. Its source prescription includes the lapse work that accompanies a change of the asymptotic AdS scale. This convention fixes the bulk charge used in the boundary calculation. The underlying Hamiltonian and covariant phase-space formulations are developed in Refs.~\cite{Regge:1974zd,Crnkovic:1986ex,Lee:1990nz,Wald:1993nt,Iyer:1994ys,Iyer:1995kg,Hollands:2005wt,Barnich:2007bf,Compere:2008us}; the cosmological-constant variation gives the extended work term \cite{Kastor:2009wy,Dolan:2011xt,Kubiznak:2016qmn,Xiao:2023lap}.

For $D=d+1\geq4$, the neutral spherical family is
\begin{equation}
 \dd s^2=-c^2 f(r)\dd t^2+\frac{\dd r^2}{f(r)}+r^2\dd\Omega_{d-1}^2,
 \qquad f(r)=1+\frac{r^2}{\ell^2}-\frac{\mu}{r^{d-2}},
 \label{eq:bulkmetric}
\end{equation}
where $\mu=r_h^{d-2}(1+r_h^2/\ell^2)$ and $r_h>0$. The dimensionless parameter $c$ labels the contraction. We define the finite Lorentzian clock and selected generator by
\begin{equation}
\begin{gathered}
 \tau=ct,\qquad s=c^\alpha t,\qquad \epsilon=c^{1-\alpha},\\[3pt]
 \xi_\lambda=\partial_s=\epsilon\partial_\tau,
 \qquad G_{d+1}(c)=c^\gamma G_C.
 \end{gathered}
 \label{eq:xilambdaG}
\end{equation}
Here $G_C>0$ and the exponents are fixed. At each $c>0$, the thermodynamic variation acts on $r_h$ and $\ell$, with $\delta c=\delta G_C=0$. Along the contraction itself we hold the geometry fixed. The adapted metric has temporal coefficient $-\epsilon^2 f$; its Carrollian branch has $\alpha<1$ and $\epsilon\to0$.

The finite-clock variables obey
\begin{equation}
\delta H_\tau=T_\tau\delta\Sth(c)+V_\tau\delta P.
 \label{eq:firstlawtau}
\end{equation}

The energy, entropy, temperature and cosmological pressure are
\begin{equation}
 \begin{aligned}
 H_\tau&=\frac{(d-1)\Omega_{d-1}\mu}{16\pi G_{d+1}},
 &\qquad \Sth(c)&=\frac{\Omega_{d-1}r_h^{d-1}}{4G_{d+1}},\\[4pt]
 T_\tau&=\frac{1}{4\pi}\left(\frac{d-2}{r_h}+\frac{d r_h}{\ell^2}\right),
 & P&=\frac{d(d-1)}{16\pi G_{d+1}\ell^2}.
 \end{aligned}
 \label{eq:parentThermoDefinitions}
\end{equation}
Their thermodynamic volume is $V_\tau=\Omega_{d-1}r_h^d/d$.

The generator change multiplies the energy, temperature and volume by $\epsilon$. Together with the coupling weight, every nonzero coefficient of the thermodynamic one-form therefore carries $c^{1-\alpha-\gamma}$. The finite branch is
\begin{equation}
 \alpha+\gamma=1,\qquad \alpha<1,
 \qquad G_{d+1}=\epsilon G_C .
 \label{eq:finitecondition}
\end{equation}
All subsequent boundary and probe limits refer to this branch. The equal weighting concerns independent variations on the equilibrium family; a special state at which one work coefficient vanishes retains the same charge normalization.

It is useful to introduce finite coordinates on that family:
\begin{equation}
\begin{aligned}
 E_C&=\epsilon H_\tau,&\qquad S_C&=\epsilon\Sth(c),&\qquad P_C&=\epsilon P,\\
 \widehat T_C&=T_\tau,& T_C&=\epsilon\widehat T_C,& V_C&=V_\tau.
 \end{aligned}
 \label{eq:finiteThermoCoordinates}
\end{equation}
Consequently, $\delta E_C=\widehat T_C\delta S_C+V_C\delta P_C$.
The entropy $\Sth(c)$ and inverse temperature $\beta_s$ of each parent theory grow as $\epsilon^{-1}$. We write $\beta_\tau=\widehat\beta_C=1/\widehat T_C$ and $\beta_s=1/T_C$. The quantities $S_C$ and $\widehat T_C$ are the normalized thermodynamic coordinates, and $E_C$ is the energy measured by the selected generator. This interpretation preserves the distinction between the Gibbs temperature and the finite coefficient used to parametrize the equilibrium family. The boundary derivation below identifies $E_C$ with a stress-tensor charge and specifies which combinations of boundary sources reproduce the $P_C$ variation. The BTZ example is treated with its three-dimensional mass and reference convention in Section~\ref{subsec:BTZ-check}.

\subsection{Stress-tensor normalization and the thermal clock}
\label{subsec:largeN-general}

The inverse Newton coupling fixes the normalization of the gravitational action and boundary stress correlations. The stress-tensor two-point coefficient $C_T$ is defined by the conformal Ward normalization \cite{Osborn:1993cr}. In Einstein holography, we write
\begin{equation}
\begin{aligned}
 C_T(c)&=\mathcal K_d\frac{\ell^{d-1}}{G_{d+1}(c)},
 &\qquad \Neff(c)&=\frac{C_T(c)}{\mathcal K_d},\\[3pt]
 \NC&=\frac{\ell^{d-1}}{G_C},
 &\Neff(c)&=\frac{\NC}{\epsilon}.
 \end{aligned}
 \label{eq:Neff}
\end{equation}
The constant $\mathcal K_d$ fixes the convention. This normalization is shared by the holographic action, stress tensor and correlation functions \cite{Maldacena:1997re,Gubser:1998bc,Witten:1998qj,Henningson:1998gx,deHaro:2000vlm,Skenderis:2002wp,Papadimitriou:2004ap,Balasubramanian:1999re,Brown:1986nw}. The symbol $\Neff$ denotes the stated gravitational normalization; a microscopic rank or central charge can differ from it by a fixed numerical factor.

On the finite branch, $T_C\Neff=\widehat T_C\NC$. Thus the generator temperature decreases as the stress-tensor normalization grows. At fixed geometry, $\Sth(c)=S_C/\epsilon$ and $\beta_s=\widehat\beta_C/\epsilon$, so the state has finite energy and finite thermal products in the chosen generator units. The large-rank limit supplies the entropy growth, while the clock transformation determines the measured temperature. Holding the finite-clock thermal scale fixed selects a sequence of equilibrium states within the parent holographic theories. The source family in Eq.~\eqref{eq:generalboundary} then determines how their normalized response coefficients are compared.

This interpretation refines the color dependence used in holographic black-hole chemistry \cite{Karch:2015rpa,Ahmed:2023snm,Frassino:2022zaz,Mann:2024sru}. The color scaling determines the leading thermodynamic factor. The present contraction also fixes the selected time generator and the weights of local source perturbations. Consequently the large-rank thermodynamic differential and the real-time Ward equations can be expressed in compatible units. Boundary pressure and generalized Euler descriptions \cite{Mancilla:2024spp,Borsboom:2026ash} remain tied to their respective boundary representatives; Sections~\ref{sec:stress} and~\ref{sec:thermodynamics} give the source representative and its thermodynamic differential.

\subsection{Type IIB normalization and independent boundary variations}
\label{subsec:AdS5CFT4}
\label{subsec:topdown-AdS5}

For type IIB theory on $\AdS_5\times S^5$, flux quantization and dimensional reduction give \cite{Maldacena:1997re,Gubser:1998bc,Witten:1998qj,Henningson:1998gx}
\begin{equation}
\begin{aligned}
 \frac{\ell^3}{G_5}&=\frac{2N^2}{\pi},&\qquad
 a=c_{4d}&=\frac{N^2}{4},\\[3pt]
 N(c)^2&=\frac{N_C^2}{\epsilon},&
 N_C^2&=\frac{\pi\ell^3}{2G_{5C}}.
 \end{aligned}
 \label{eq:AdS5N2}
\end{equation}
The displayed central charges are the leading large-rank terms. At fixed large 't Hooft coupling, $g_s=\lambda/(4\pi N)$ tends to zero along an integer-rank sequence. Ref.~\cite{Xu:2026vpj} gives the associated supergravity normalization and thermal comparison. We use that sequence here to identify the boundary stress factor, with $\Neff=2N^2/\pi$. The equilibrium gravitational calculation is at leading order in inverse rank and in the large-coupling expansion. Section~\ref{sec:regge} examines the additional coupling hierarchy selected by an interacting observable on the logarithmic scrambling interval. The construction of a contracted microscopic string and gauge theory in Ref.~\cite{Fontanella:2025tbs} addresses the corresponding microscopic dynamics through its own limiting source prescription.

The boundary thermodynamic variation extends the fixed-radius comparison by allowing independent spatial volume and holographic normalization. With $\mathcal V_\partial=\Omega_3\ell^3$ and $\NC=\ell^3/G_{5C}$, their logarithmic variations are
\begin{equation}
 \delta\log\mathcal V_\partial=3\,\delta\log\ell,
 \qquad
 \delta\log\NC=3\,\delta\log\ell-\delta\log G_{5C}.
 \label{eq:independentBoundaryVariations}
\end{equation}
At fixed $G_{5C}$ the two variations coincide. At fixed boundary volume a variation of $\NC$ changes the coupling normalization, and in the type IIB interpretation it changes the leading color factor. The partial derivatives used to define the boundary pressure and color chemical potential therefore belong to this enlarged thermodynamic parameter space. The bulk pressure differential at fixed $G_{5C}$ is recovered by restricting to its prescribed curve after those derivatives have been computed. This order distinguishes a variation of the source radius from a variation of the normalization at fixed radius.

An integer microscopic rank becomes a continuous thermodynamic parameter at leading large rank: adjacent values change $N$ by one and $N^2$ by $2N+1$, so their fractional separation tends to zero. The derivative with respect to $\NC$ refers to this leading equation of state. The clock factor $\epsilon$ is held fixed during the derivative. Motion along the contraction sequence instead changes the parent rank and thermal period together while retaining the finite coordinates. These two comparisons yield different derivatives and are used for different purposes below.

\subsection{BTZ mass, entropy and central charge}
\label{subsec:AdS3CFT2}
\label{subsec:BTZ-check}
\label{subsec:anomaly-cardy}

The nonrotating BTZ family provides a lower-dimensional normalization check. With $f=(r^2-r_+^2)/\ell^2$, its finite-clock mass relative to massless BTZ, temperature and entropy are
\begin{equation}
\begin{aligned}
 M_\tau&=\frac{r_+^2}{8G_3\ell^2},&\qquad T_\tau&=\frac{r_+}{2\pi\ell^2},\\[3pt]
 S&=\frac{\pi r_+}{2G_3},&c_{\CFT}&=\frac{3\ell}{2G_3}.
 \end{aligned}
 \label{eq:BTZthermo}
\end{equation}
The Brown--Henneaux central charge fixes the normalization \cite{Brown:1986nw}. With $G_3=\epsilon G_{3C}$, the selected-clock mass is finite, and both $S$ and $c_{\CFT}$ carry $\epsilon^{-1}$. The reference to global AdS adds $1/(8G_{3C})$ to the selected-clock BTZ mass. We retain this shift whenever comparing the two thermal geometries. At fixed $\ell$ and coupling it leaves variations with respect to $r_+$ unchanged; a variation of the normalization must also differentiate the shift. The distinction makes the BTZ reference consistent with the common-source subtraction used for the higher-dimensional neutral family.

In the large-central-charge black-hole regime, the thermal Cardy expression is
\begin{equation}
 S_{\rm BTZ}=\frac{2\pi^2\ell}{3}c_{\CFT}T_\tau.
 \label{eq:BTZCardyThermal}
\end{equation}
Substitution of Eq.~\eqref{eq:BTZthermo} reproduces the horizon entropy. Keeping $T_\tau\ell$ fixed above the dominant-black-hole threshold selects the corresponding high-energy holographic saddle. The selected temperature is $T_C=\epsilon T_\tau$, so $T_C\delta S$ remains finite. In particular, global AdS and BTZ exchange dominance at $r_+=\ell$, where the reference energy and thermal entropy contributions have matching magnitudes. This check uses the established saddle interpretation and fixes the thermal normalization against a known boundary density of states.

The Weyl anomaly has the same central-charge factor as the stress tensor. On the static flat cylinder its local curvature contribution vanishes, while the Casimir energy retains its prescribed reference value. For curved sources, extracting the common central-charge factor yields a finite anomaly coefficient together with the normalized stress response. The source variation and reference subtraction must therefore be performed in the same convention when the boundary geometry changes. This is the lower-dimensional counterpart of the source dependence retained in Sections~\ref{sec:stress} and~\ref{sec:thermodynamics}. The relation to BMS symmetry and conformal Carroll contractions is developed in Refs.~\cite{PhysRevLett.105.171601,Bagchi:2012cy,Ciambelli:2018wre,Campoleoni:2018ltl,Bagchi:2016bcd,Dutta:2022vkg,Have:2024dff}.

\subsection{Normalization of finite responses}
\label{subsec:thermal-summary}

Table~\ref{tab:dictionary} collects the normalization map used in the rest of the paper. Each finite coefficient is obtained by specifying both a parent observable and its source or clock weight. The energy is a selected-generator charge. The entropy coefficient parametrizes the leading density of states. The scalar response includes the operator source map derived in Section~\ref{sec:scalar}; the same map fixes the normalization of the interacting four-point observable in Section~\ref{sec:regge}. These assignments explain why quantities that share the same large-rank factor can acquire different frequency weights after a change of time coordinate.

\begin{table}[tb]
 \centering
 \renewcommand{\arraystretch}{1.2}
 \begin{tabular}{lll}
 \toprule
 Parent quantity & Scaling & Finite boundary quantity\\
 \midrule
 $C_T$ & $\epsilon^{-1}$ & $\epsilon C_T=\mathcal K_d\NC$\\
 $N^2$ in the type IIB example & $\epsilon^{-1}$ & $\epsilon N^2=N_C^2$\\
 $c_{\CFT}$ for BTZ & $\epsilon^{-1}$ & $3\ell/(2G_{3C})$\\
 $T_C$ & $\epsilon$ & $\widehat T_C=T_C/\epsilon$\\
 $\Sth(c)$ & $\epsilon^{-1}$ & $S_C=\epsilon\Sth(c)$\\
 $H_\tau$ & $\epsilon^{-1}$ & $E_C=\epsilon H_\tau$\\
 $T_{\tau A}$ & $\epsilon^{-1}$ & $\mathcal P_A^C$\\
 $\Phi_\tau$ source coupling & $\dd\tau=\epsilon\dd s$ & $\Phi_s=\epsilon\Phi_\tau(\epsilon s)$\\
 \bottomrule
 \end{tabular}
 \caption{Boundary normalization map on the finite Carroll branch. The rank factors depend on the microscopic realization, while the source and generator assignments fix the stress and probe observables.}
 \label{tab:dictionary}
\end{table}

\section{Boundary sources and stress charges}
\label{sec:boundary}
\label{sec:stress}

The normalization of Section~\ref{sec:bulk} becomes a boundary observable through the metric sources and their conjugate stress tensor. We specify the lapse, shift and spatial metric before evaluating the homogeneous state. The resulting Brown--York energy then fixes the charge used in the thermodynamic and local conservation identities.

\subsection{Clock, shift and spatial sources}

The conformal representative induced by the neutral AdS family is
\begin{equation}
 \dd s^2_{\partial,c}=-c^2\dd t^2+\ell^2q_{AB}\dd x^A\dd x^B
 =-\epsilon^2\dd s^2+\ell^2q_{AB}\dd x^A\dd x^B .
 \label{eq:boundarymetric}
\end{equation}
At finite $c$ this metric defines a Lorentzian cylinder. In the selected clock $s$, its limiting Carroll data are the clock form $\vartheta=\dd s$, spatial metric $h_{AB}=\ell^2q_{AB}$ and temporal vector $v=\partial_s$. They obey $\vartheta_\mu v^\mu=1$ and $h_{\mu\nu}v^\nu=0$. The inverse spatial metric is defined on the spatial subspace, with $h^{AB}=\ell^{-2}q^{AB}$ and $h^{\mu\nu}\vartheta_\nu=0$. We use $\vartheta$ for this one-form and reserve $\tau$ for the finite Lorentzian coordinate.

Momentum response depends on the way an off-diagonal metric perturbation approaches the cylinder. We use the family
\begin{equation}
 \dd s^2_{\partial,\epsilon}
 =-\epsilon^2\Omega^2\dd s^2
 +h_{AB}(\dd x^A+\epsilon B^A\dd s)
          (\dd x^B+\epsilon B^B\dd s),
 \label{eq:generalboundary}
\end{equation}
where the lapse $\Omega$, scaled shift $B^A$ and spatial metric $h_{AB}$ are varied independently before specializing to $\Omega=1$, $B^A=0$ and the round sphere. The factor $\epsilon$ assigns the shift a weight compatible with finite momentum response. A perturbation of $B^A$ changes the finite-clock shift by an order-one amount; its coordinate component in the $s$ metric has the factor required by the contraction. The coframe is $e^0=\epsilon\Omega\dd s$ and $e^a=E^a{}_A(\dd x^A+\epsilon B^A\dd s)$, with $h_{AB}=\delta_{ab}E^a{}_AE^b{}_B$.

This family retains the distinction between momentum density and energy flux. At finite $\epsilon$, the symmetric relativistic stress tensor relates them through the metric. The relation contains powers of the clock parameter, so its limiting components have different roles in a conservation equation. Varying Eq.~\eqref{eq:generalboundary} determines those powers from the source coupling. Section~\ref{sec:ward} derives the resulting response functional and shows how spatial transport enters at the thermal time scale.

The asymptotic representative is also part of the thermodynamic prescription. At fixed $\ell$, source-dependent counterterms cancel between the black-hole and AdS contributions to the stress tensor. When $\ell$ varies, we compare the two geometries at the corresponding common source and retain its variation. Thus the subtraction fixes a family of energy zeros together with the source geometry. This agrees with the static boundary prescription in Ref.~\cite{Xu:2026vpj}. On the boundary it induces a variation of the sphere volume and holographic normalization. The same source data therefore enter both the local response problem and the global thermodynamic differential.

On a finite interval of $s$, the bulk action density has a temporal Jacobian $\epsilon$ and an inverse coupling proportional to $\epsilon^{-1}$. Their product is finite on Eq.~\eqref{eq:finitecondition}. On a full thermal circle the interval itself grows as $\epsilon^{-1}$, producing the large-rank thermal saddle exponent. The local response and the Gibbs saddle thus arise from the same renormalized functional evaluated on different time domains. The local Ward equations concern source variations on finite intervals, whereas the partition function in Section~\ref{sec:partition} uses the full KMS period.

\subsection{Renormalized stress response and the finite charge}
\label{subsec:contracted-BY-charge}

The boundary realization of the contracted Hamiltonian follows from the renormalized Brown--York tensor. We use $\delta I_{\ren}=\frac12\int\sqrt{-\gamma}\,T^{ij}\delta\gamma_{ij}$, so the lower-index tensor is $T_{ij}=-2(-\gamma)^{-1/2}\delta I_{\ren}/\delta\gamma^{ij}$. At a radial cutoff with induced metric $\gamma_{ij}$ and outward normal $n$, choose $K_{ij}=-\frac12\mathcal L_n\gamma_{ij}$. For $d>2$, the counterterm expression is \cite{Balasubramanian:1999re,deHaro:2000vlm,Skenderis:2002wp}
\begin{equation}
 T^{\BY}_{ij}
 =\frac{1}{8\pi G_{d+1}}
 \left[K_{ij}-K\gamma_{ij}-\frac{d-1}{\ell}\gamma_{ij}
       +\frac{\ell}{d-2}\mathcal G_{ij}[\gamma]+\cdots\right].
 \label{eq:BYtensor}
\end{equation}
The additional local terms are the counterterms appropriate to the boundary dimension. For $d=2$ the local expression contains the extrinsic-curvature and volume terms, supplemented by the logarithmic anomaly prescription. This case has the BTZ reference convention stated in Section~\ref{subsec:BTZ-check}. The remaining stress calculation in this section uses $d\geq3$.

We subtract AdS at the same boundary source and denote the difference by $\Delta T^{\BY}_{ij}$. The stress tensor in the boundary representative is obtained by multiplying the cutoff difference by $(R/\ell)^{d-2}$ before taking $R\to\infty$. Counterterms depending only on that common source cancel in the difference. At a fixed source, a source-dependent vacuum energy leaves the state variation unchanged. Under an extended variation of $\ell$, its derivative would shift the thermodynamic conjugates. Our convention assigns zero energy to the corresponding AdS reference for each $\ell$ and includes the resulting source work, as prescribed in Ref.~\cite{Xu:2026vpj}. The subtraction thus fixes the complete extended energy function.

In the finite Lorentzian clock, the subtracted state is homogeneous on the sphere. Its stress components are
\begin{equation}
\begin{aligned}
 \Delta T^{\BY}_{\tau\tau}&=\frac{(d-1)\mu}{16\pi G_{d+1}\ell^{d-1}},
 &\qquad \Delta T^{\BY}_{\tau A}&=0,\\[3pt]
 \Delta T^{\BY}_{AB}&=\frac{\mu}{16\pi G_{d+1}\ell^{d-3}}q_{AB}.
 \end{aligned}
 \label{eq:BYfluid}
\end{equation}
The pressure is the energy density divided by $d-1$, and integration over the sphere gives $H_\tau$ in Eq.~\eqref{eq:parentThermoDefinitions}. The trace of the subtracted tensor vanishes: the source-dependent anomalous part has canceled against the common reference. Its state-dependent part therefore has the conformal perfect-fluid form.

The finite stress variables are selected by the frame weights and the coupling normalization:
\begin{equation}
 \begin{aligned}
 \mathcal E_C&=\lim_{c\to0}c^{\gamma-2}\Delta T^{\BY}_{tt}
             =\lim_{\epsilon\to0}\epsilon\Delta T^{\BY}_{\tau\tau},\\
 \mathcal P_A^C&=\lim_{c\to0}c^{\gamma-1}\Delta T^{\BY}_{tA}
             =\lim_{\epsilon\to0}\epsilon\Delta T^{\BY}_{\tau A},\\
 \mathcal T_{AB}^C&=\lim_{c\to0}c^\gamma\Delta T^{\BY}_{AB}.
 \end{aligned}
 \label{eq:finiteBY}
\end{equation}
The factors in the $t$ coordinate arise from $\tau=ct$; the equivalent expressions in the finite-clock frame use $c^\gamma=\epsilon$ on the selected branch. For the neutral black hole,
\begin{equation}
\begin{aligned}
 \mathcal E_C&=\frac{(d-1)\mu}{16\pi G_C\ell^{d-1}},&\qquad\mathcal P_A^C&=0,\\[3pt]
 \mathcal T_{AB}^C&=p_C h_{AB},&p_C&=\frac{\mathcal E_C}{d-1}.
 \end{aligned}
 \label{eq:finiteBYresult}
\end{equation}
Here $p_C$ is the boundary fluid pressure, conjugate to the spatial source volume. The uppercase $P_C$ of Eq.~\eqref{eq:finiteThermoCoordinates} is the normalized cosmological pressure, whose conjugate is the bulk thermodynamic volume. Section~\ref{sec:thermodynamics} derives the relation between these work terms.

The selected-generator Brown--York charge has the same weight. A unit normal to the boundary time slice is $u=\epsilon^{-1}\partial_s$, and $\xi_\lambda=\partial_s$. Hence
\begin{equation}
\begin{aligned}
 Q[\xi_\lambda]
 &=\int\dd^{d-1}x\sqrt h\,\epsilon^{-1}\Delta T^{\BY}_{ss}
 \longrightarrow E_C,\\
 E_C&=\int\dd^{d-1}x\sqrt h\,\mathcal E_C
 =\frac{(d-1)\Omega_{d-1}\mu}{16\pi G_C}.
 \end{aligned}
 \label{eq:EC}
\end{equation}
The resulting charge equals the selected-generator bulk energy in Eq.~\eqref{eq:finiteThermoCoordinates}. Its derivation from the tensor response fixes both the integration measure and the temporal frame. The same energy component consequently enters the boundary equation of state and the local source variation.

For independent variations of the horizon and AdS radii,
\begin{equation}
 \delta E_C=\frac{(d-1)\Omega_{d-1}}{16\pi G_C}
 \left[\left((d-2)r_h^{d-3}+\frac{d r_h^{d-1}}{\ell^2}\right)\delta r_h
       -\frac{2r_h^d}{\ell^3}\delta\ell\right].
 \label{eq:deltaEC}
\end{equation}
The horizon coefficient agrees with $\widehat T_C\delta S_C$, and the AdS-radius coefficient agrees with $V_C\delta P_C$. Section~\ref{sec:thermodynamics} resolves the latter coefficient into boundary source variations. This identifies how the same extended work is measured by the boundary fluid and by its holographic normalization.

\section{Boundary thermodynamics and the thermal ensemble}
\label{sec:thermodynamics}
\label{sec:partition}

The finite stress charge defines an energy function on the space of boundary states and sources. Its derivatives determine the temperature, spatial pressure and chemical potential for the holographic normalization. This section connects those derivatives to the Gibbs ensemble, its leading energy fluctuations and the Hawking--Page saddle crossing. The same ensemble supplies the states used for the scalar and four-point calculations.

\subsection{Volume and normalization work}
\label{subsec:boundary-first-law}

To resolve the AdS-radius work into boundary variables, introduce the spatial source volume and dimensionless horizon radius,
\begin{equation}
        \mathcal V_{\partial}=\Omega_{d-1}\ell^{d-1},
        \qquad
        x=\frac{r_h}{\ell} .
        \label{eq:boundaryVariables}
\end{equation}
The finite energy and normalized entropy coordinate can be written as
\begin{equation}
\begin{aligned}
 E_C&=\frac{(d-1)\Omega_{d-1}}{16\pi}\frac{\NC}{\ell}
        x^{d-2}(1+x^2),\\
 S_C&=\frac{\Omega_{d-1}}{4}\NC x^{d-1}.
 \end{aligned}
 \label{eq:boundaryEnergyEntropyVariables}
\end{equation}
These expressions define the boundary energy function $E_C(S_C,\mathcal V_\partial,\NC)$. Partial derivatives are taken at fixed contraction parameter with the remaining boundary coordinates independent. The resulting differential is
\begin{equation}
        \delta E_C
        =
        \widehat T_C\delta S_C
        -p_{\partial}\delta \mathcal V_{\partial}
        +\mu_{\NC}\delta\NC .
        \label{eq:boundaryFirstLawVariables}
\end{equation}
The temperature conjugate to the boundary entropy variable is
\begin{equation}
\begin{aligned}
 \widehat T_C&:=\left(\frac{\partial E_C}{\partial S_C}\right)_{\mathcal V_{\partial},\NC}\\
 &=\frac{1}{4\pi\ell}\left(\frac{d-2}{x}+d x\right).
 \end{aligned}
 \label{eq:boundaryThermalTemperature}
\end{equation}
This conjugate equals the finite-clock temperature of the thermal ensemble. At fixed $S_C$ and $\NC$, the boundary pressure is
\begin{equation}
        p_{\partial}
        :=
        -\left(\frac{\partial E_C}{\partial \mathcal V_{\partial}}\right)_{S_C,\NC}
        =
        \frac{E_C}{(d-1)\mathcal V_{\partial}} .
        \label{eq:boundaryPressure}
\end{equation}
The result agrees with the homogeneous stress coefficient $p_C$ in Eq.~\eqref{eq:finiteBYresult}. We use $p_\partial$ when treating this pressure as a derivative on the enlarged thermodynamic state space. This agreement checks that varying the source volume and integrating the stress tensor define the same mechanical response.
Varying $\NC$ at fixed $S_C$ and $\mathcal V_\partial$ changes both the overall normalization and the entropy per degree of freedom. The conjugate chemical potential is
\begin{equation}
\begin{aligned}
 \mu_{\NC}&:=\left(\frac{\partial E_C}{\partial\NC}\right)_{S_C,\mathcal V_{\partial}}\\
 &=\frac{\Omega_{d-1}}{16\pi\ell}\left(x^{d-2}-x^d\right).
 \end{aligned}
 \label{eq:boundaryChemicalPotential}
\end{equation}
The bulk comparison selects a curve in this enlarged state space. Varying $\ell$ at fixed $r_h$ and $G_C$ keeps $S_C$ fixed and gives
\begin{equation}
        \frac{\delta \mathcal V_{\partial}}{\mathcal V_{\partial}}
        =
        (d-1)\frac{\delta\ell}{\ell},
        \qquad
        \frac{\delta\NC}{\NC}
        =
        (d-1)\frac{\delta\ell}{\ell} .
        \label{eq:boundaryVariationsEll}
\end{equation}
The mechanical and normalization contributions must be evaluated along this same curve. Substituting Eq.~\eqref{eq:boundaryPressure} and Eq.~\eqref{eq:boundaryChemicalPotential} gives
\begin{equation}
        -p_{\partial}\delta \mathcal V_{\partial}
        +
        \mu_{\NC}\delta\NC
        =
        -\frac{(d-1)\Omega_{d-1}}{8\pi G_C}
        \frac{r_h^d}{\ell^3}\delta\ell .
        \label{eq:boundaryPressureChemicalCombination}
\end{equation}
Comparing this expression with the finite bulk pressure contribution gives
\begin{equation}
        V_C\delta P_C
        =
        -p_{\partial}\delta \mathcal V_{\partial}
        +
        \mu_{\NC}\delta\NC .
        \label{eq:PVBoundaryIdentity}
\end{equation}
Eq.~\eqref{eq:PVBoundaryIdentity} identifies the boundary sources whose combined variation reproduces the extended bulk work. Holding the boundary volume fixed while changing only the color factor would follow a different curve and yield a different conjugate. The common-source prescription specifies the curve appropriate to the static Hamiltonian contraction. This implements the holographic black-hole-chemistry description of the AdS-scale variation in finite boundary coordinates \cite{Karch:2015rpa,Johnson:2014yja,Ahmed:2023snm,Frassino:2022zaz,Mann:2024sru}. The spatial pressure can also be compared with quasilocal constructions that define thermodynamic variables directly from boundary data \cite{Borsboom:2026ash}. The temperature, volume and normalization terms here are derivatives of a single energy function; Appendix~\ref{app:conventions} develops its mixed susceptibilities and fluctuation directions.

The same boundary variables give a thermodynamic interpretation of the Hawking--Page locus.  Under a common scaling $(S_C,\mathcal V_\partial,\NC)\mapsto a(S_C,\mathcal V_\partial,\NC)$, the ratio $x$ is fixed and $\ell\mapsto a^{1/(d-1)}\ell$. The energy therefore has degree $(d-2)/(d-1)$ under this scaling. Differentiating with respect to $a$ gives the weighted Euler relation
\begin{equation}
        \frac{d-2}{d-1}E_C
        =
        \widehat T_C S_C
        -
        p_\partial\mathcal V_\partial
        +
        \mu_{\NC}\NC .
        \label{eq:boundaryEulerRelation}
\end{equation}
Using \(p_\partial\mathcal V_\partial=E_C/(d-1)\), Eq.~\eqref{eq:boundaryEulerRelation} becomes
\begin{equation}
        F_C
        :=
        E_C-\widehat T_C S_C
        =
        \NC\mu_{\NC}.
        \label{eq:freeEnergyChemicalPotential}
\end{equation}
The finite free energy is therefore the product of the holographic normalization and its conjugate chemical potential. Its homogeneity follows from the boundary curvature radius as well as the color factor. Fluid-action descriptions similarly organize the Smarr identity through a generalized Euler relation with an additional geometric scale \cite{Mancilla:2024spp}. Here that scale is the source radius, whose variation is correlated with the normalization by the contraction. The degree under a common scaling of all boundary coordinates differs from the degree under scaling only $S_C$ and $\NC$ at fixed volume. The latter leaves $x$ unchanged and makes the energy linear in the common normalization. These two scaling operations account for the Euler coefficient and the null susceptibility direction derived in Appendix~\ref{app:conventions}. From Eq.~\eqref{eq:boundaryChemicalPotential}, the Hawking--Page radius $x=1$ is the zero of $\mu_{\NC}$. The large-black-hole branch has $\mu_{\NC}<0$ and negative free energy for $x>1$; for $x<1$, both quantities are positive relative to the thermal AdS reference.

Combining Eq.~\eqref{eq:boundaryFirstLawVariables} and Eq.~\eqref{eq:PVBoundaryIdentity} gives the extended thermodynamic identity as a boundary stress-charge variation. The Gibbs construction below uses this energy function and the same fixed-source comparison to define the thermal state.

\subsection{Density of states and the large-rank exponent}
\label{subsec:ensemble}
\label{subsec:double-scaled-ensemble}

At each nonzero $\epsilon$, let $\mathcal H_\epsilon$ be the Hilbert space of the parent boundary theory and $H_{s,\epsilon}=\epsilon H_{\tau,\epsilon}$ its selected-generator Hamiltonian. The Gibbs state is
\begin{equation}
 \rho_\epsilon=Z_\epsilon^{-1}e^{-\beta_sH_{s,\epsilon}},\qquad
 \beta_s=\frac{\widehat\beta_C}{\epsilon},\qquad
 \beta_sH_{s,\epsilon}=\widehat\beta_C H_{\tau,\epsilon}.
 \label{eq:rho}
\end{equation}
The thermal weights agree under a change of clock within each parent theory. Moving along the large-rank sequence also changes the density of states. The source and observable maps specify how those states are compared, while $\rho_\epsilon$ defines the expectation value at every nonzero $\epsilon$. A limiting correlator is therefore constructed from mapped observables and a specified observation interval. The scalar spectral response in Section~\ref{sec:scalar} and the normalized four-point functional in Section~\ref{sec:regge} provide complementary constructions on the thermal and scrambling intervals.

For selected-generator energy $E=O(1)$, the black-hole density of states has the leading semiclassical entropy $\mathscr S_\epsilon(E)=\epsilon^{-1}S_C(E)+O(1)$ within a dominant phase. Its contribution to the partition function takes the form
\begin{equation}
 Z_\epsilon\sim\int\dd E\,
 \exp\!\left[\frac{S_C(E)-\widehat\beta_C E}{\epsilon}+O(1)\right].
 \label{eq:saddle}
\end{equation}
The stationary condition is $\partial_ES_C=\widehat\beta_C$. Thus the entropy and Boltzmann terms remain of the same order in the exponent. The boundary normalization is $\Neff=\NC/\epsilon$, and the saddle action per unit holographic normalization is
\begin{equation}
 \widehat I_C:=\frac{I_\epsilon}{\Neff(c)}
 =\widehat\beta_C\frac{F_C}{\NC}.
 \label{eq:actionDensity}
\end{equation}
The finite potential $F_C$ gives the leading thermal free energy in the selected-generator units. The entropy coordinate $S_C$ is the leading entropy divided by its growing normalization. These assignments preserve the thermodynamic pairing and distinguish the finite saddle coefficient from the entropy of an individual finite-rank theory.

There is a useful consequence for the global stress response within a stable black-hole phase. Let $C_C=(\partial E_C/\partial\widehat T_C)_{\mathcal V_\partial,\NC}$ be the normalized heat capacity. Differentiating the Gibbs weight at fixed theory gives
\begin{equation}
 \operatorname{Var}_{\rho_\epsilon}(H_{s,\epsilon})
 =\epsilon\widehat T_C^2C_C+o(\epsilon).
 \label{eq:boundaryEnergyVariance}
\end{equation}
The finite energy therefore has fluctuations of order $\sqrt\epsilon$ around its saddle value. The variance sets the zero-frequency weight of the connected Brown--York energy correlation. The factor also agrees with the connected correlator of a source-normalized single-trace operator: two operator weights supply $\epsilon^2$, while the parent connected two-point normalization supplies $\epsilon^{-1}$. The time dependence is resolved on $\tau=\epsilon s$, as required by the Ward equations. On the longer scrambling interval, the horizon boost can compensate this rank suppression in a normalized four-point observable; Section~\ref{sec:regge} determines the coupling conditions for that limit. At a phase crossing the full Gibbs state can contain phase-mixture fluctuations; Eq.~\eqref{eq:boundaryEnergyVariance} refers to a selected stable branch. Ref.~\cite{Xu:2026vpj} develops the reduced gravitational ensemble and its phase weights.

\subsection{KMS condition in the selected clock}

We use Heisenberg evolution $\Phi_s(s)=e^{iH_{s,\epsilon}s}\Phi_s(0)e^{-iH_{s,\epsilon}s}$, so $[H_{s,\epsilon},\Phi_s]=-i\partial_s\Phi_s$. Thermal cyclicity gives \cite{Kubo:1957mj,Martin:1959jp,Haag:1967sg}
\begin{equation}
 \left\langle\Phi_1(s-i\beta_s,x_1)\Phi_2(0,x_2)\right\rangle_{\rho_\epsilon}
 =\left\langle\Phi_2(0,x_2)\Phi_1(s,x_1)\right\rangle_{\rho_\epsilon}.
 \label{eq:CarrollianKMS}
\end{equation}
The imaginary-time interval and Hamiltonian use the same generator. With $\tau=\epsilon s$, the thermal circle has fixed finite-clock circumference $\widehat\beta_C$ and growing selected-clock circumference $\beta_s$. A mode of frequency $\omega=\epsilon\Omega$ then has the invariant occupation argument $\beta_s\omega=\widehat\beta_C\Omega$. This is the interval probed by the spectral and Mellin observables below. The vanishing generator temperature is therefore compatible with nontrivial KMS dependence after the frequency window is transformed with the clock.

In the static black-hole state the angular stress profile is homogeneous, and the equilibrium density matrix is invariant under the global time generator. Nonconstant time shifts act on the sources according to Section~\ref{sec:ward}. Their insertion identity includes that source response, so finite-frequency angular correlators retain their time dependence consistently with stationarity. This connects the density matrix, the local Ward equation and the frequency-diagonal two-point kernel within one normalization convention.

\subsection{Boundary chemical potential and the Hawking--Page crossing}
\label{subsec:freeenergyHP}

With the AdS reference and source prescription of Section~\ref{sec:stress}, the neutral free energy is the established Schwarzschild-AdS saddle potential \cite{Hawking:1982dh,Witten:1998zw} expressed in finite coordinates:
\begin{equation}
 F_C=\frac{\Omega_{d-1}r_h^{d-2}}{16\pi G_C}
       \left(1-\frac{r_h^2}{\ell^2}\right).
 \label{eq:FC}
\end{equation}
The crossing is at $r_h=\ell$ and $\widehat T_{C,\mathrm{HP}}=(d-1)/(2\pi\ell)$. The selected-clock transition temperature is $\epsilon\widehat T_{C,\mathrm{HP}}$. The geometric location is inherited from the parent ensemble, while the boundary description identifies its finite color chemical potential and stress response. This comparison uses the leading thermodynamic potentials of the competing saddles.

\begin{figure}[tb]
 \centering
 \includegraphics[width=0.49\linewidth]{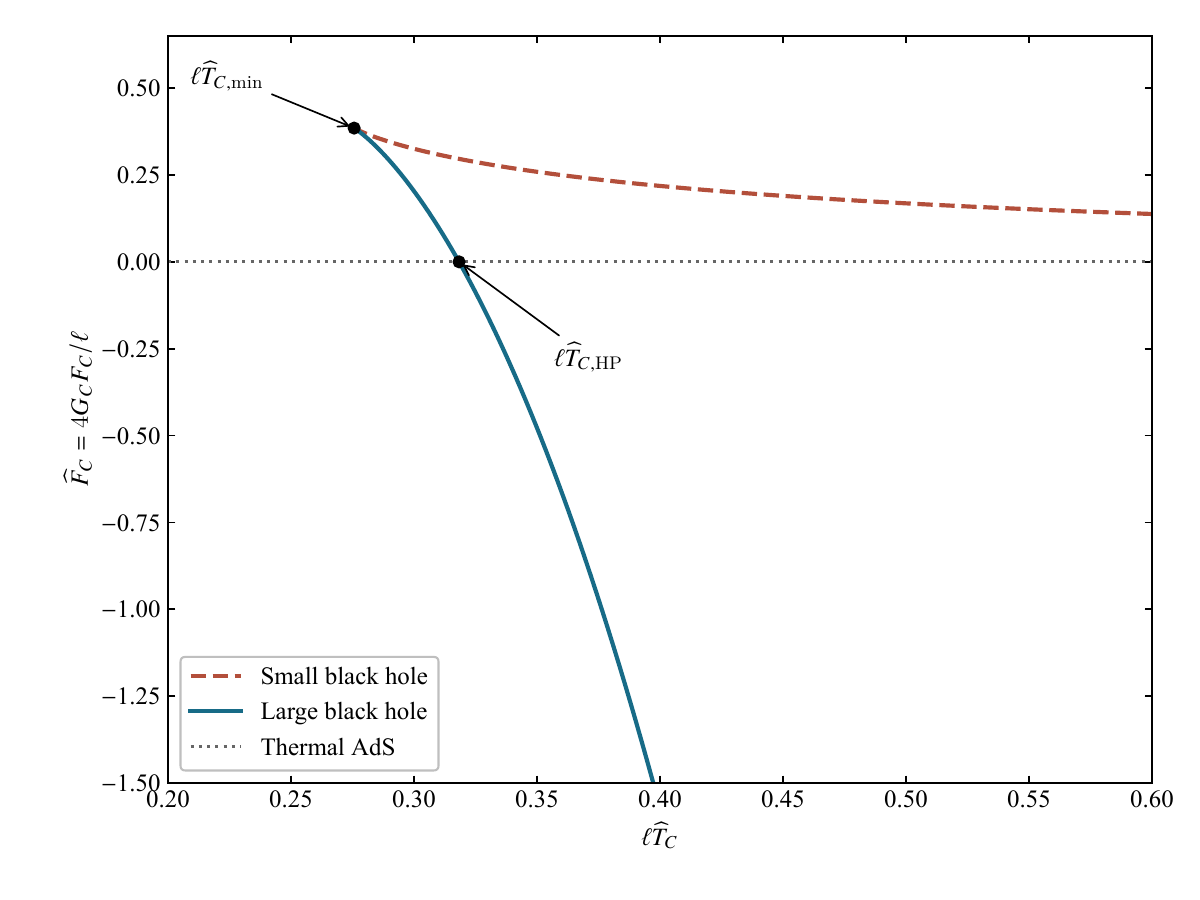}
 \caption{Boundary saddle potential of the four-dimensional Schwarzschild-AdS family. The temperature and free energy are normalized as in Eq.~\eqref{eq:dimensionlessHP}. The zero of the potential is also the zero of the holographic chemical potential. The large-black-hole branch supplies the stable states used for the scalar response calculation.}
 \label{fig:carrollianHP}
\end{figure}

For $d=3$, define $x=r_h/\ell$ and $\widehat F_C=4G_CF_C/\ell$. The curve in Fig.~\ref{fig:carrollianHP} is parametrized by
\begin{equation}
 \ell\widehat T_C=\frac{1+3x^2}{4\pi x},\qquad
 \widehat F_C=x(1-x^2).
 \label{eq:dimensionlessHP}
\end{equation}
The branch merger is at $x=1/\sqrt3$, where $\ell\widehat T_C=\sqrt3/(2\pi)$; the thermal crossing is at $x=1$, where $\ell\widehat T_C=1/\pi$. Numerically these temperatures are $0.2757$ and $0.3183$. The stable branch between them has positive free energy relative to AdS. Above the crossing, the large-black-hole branch has the lower free energy and a negative holographic chemical potential.

For the boundary heat capacity at fixed $\ell$ and $\NC$, differentiation of the same equation of state gives
\begin{equation}
 C_C=(d-1)S_C\,
 \frac{d x^2+d-2}{d x^2-(d-2)}.
 \label{eq:boundaryHeatCapacity}
\end{equation}
Its sign agrees with the local stability of the black-hole branch, and its divergence marks the branch merger. Together with Eq.~\eqref{eq:boundaryEnergyVariance}, it fixes the leading integrated energy fluctuations in the boundary ensemble. The scalar response in Section~\ref{sec:scalar} probes complementary information: its angular coefficients distinguish horizons of different radii even when their energy and temperature have already been specified. The equilibrium potential selects the thermal state, while the source-dependent radial response determines its absorptive observables.

\section{Sourced Ward identities and time-shift charges}
\label{sec:ward}

The static stress tensor fixes an equilibrium charge. To determine the local conservation law we keep the source dependence of the parent functional before taking the limit. Let $W_\epsilon[g]$ denote the renormalized Lorentzian functional with the AdS reference subtracted at the same source. We use the convention $\delta W_\epsilon=\frac12\int\sqrt{-g}\,T^{ij}\delta g_{ij}$. At the static source in Eq.~\eqref{eq:generalboundary}, define finite-$\epsilon$ response coefficients by
\begin{equation}
 \mathcal E_\epsilon=\epsilon T_{\tau\tau},\qquad
 \mathcal P_{\epsilon A}=\epsilon T_{\tau A},\qquad
 \mathcal T_{\epsilon AB}=\epsilon T_{AB}.
 \label{eq:finiteResponseCoefficients}
\end{equation}
These definitions apply to perturbations of the subtracted state and coincide in the limit with Eq.~\eqref{eq:finiteBY}. Coordinate components obey $T_{ss}=\epsilon\mathcal E_\epsilon$ and $T_{sA}=\mathcal P_{\epsilon A}$. We raise spatial indices with $h^{AB}$ and write $D_A$ for its covariant derivative.

Varying the lapse, shift and spatial metric independently gives
\begin{equation}
 \left.\delta W_\epsilon\right|_{\Omega=1,B=0}
 =\int\dd s\,\dd^{d-1}x\sqrt h\left[
 -\mathcal E_\epsilon\delta\Omega
 -\mathcal P_{\epsilon A}\delta B^A
 +\frac12\mathcal T_\epsilon^{AB}\delta h_{AB}\right].
 \label{eq:sourcedVariation}
\end{equation}
The response to $B^A$ carries the sign of the covariant time-space stress component in our convention. This sign follows by raising the temporal index with $g^{ss}=-\epsilon^{-2}$ at $\Omega=1$. The temporal measure $\epsilon\sqrt h$ at this source cancels the inverse-coupling normalization. Eq.~\eqref{eq:sourcedVariation} therefore has finite coefficients on the static family and on source perturbations whose normalized stress responses remain bounded. The limiting response is determined by this variational pairing, including its shift normalization.

The parent diffeomorphism identity is $\nabla_iT^{ij}=0$. At a static product source its covariant-component form is
\begin{equation}
 -\epsilon^{-2}\partial_sT_{ss}+D^AT_{As}=0,
 \qquad
 -\epsilon^{-2}\partial_sT_{sA}+D^BT_{BA}=0.
 \label{eq:relward}
\end{equation}
Substituting Eq.~\eqref{eq:finiteResponseCoefficients} yields
\begin{align}
 \partial_s\mathcal E_\epsilon-\epsilon D_A\mathcal P_\epsilon^A&=0,
 \label{eq:wardE}\\
 \partial_s\mathcal P_{\epsilon A}-\epsilon D_B\mathcal T_\epsilon^B{}_A&=0.
 \label{eq:wardP}
\end{align}
The same factors follow directly from Eq.~\eqref{eq:sourcedVariation}. An infinitesimal time diffeomorphism $f(s,x)\partial_s$ induces $\delta\Omega=\partial_sf$ and $\delta B^A=-\epsilon D^Af$ at the static source. Inserting these variations and integrating by parts gives Eq.~\eqref{eq:wardE}. A spatial diffeomorphism with parameter $\zeta^A(s,x)$ induces $\delta B^A=\epsilon^{-1}\partial_s\zeta^A$ and $\delta h_{AB}=2D_{(A}\zeta_{B)}$, which gives Eq.~\eqref{eq:wardP}. Compactly supported parameters remove endpoint terms. This derivation tracks the clock and shift contributions within the same functional.

In the selected time $s$, the normalized energy flux is $\mathcal J_\epsilon^A=-\epsilon\mathcal P_\epsilon^A$. The energy equation is thus $\partial_s\mathcal E_\epsilon+D_A\mathcal J_\epsilon^A=0$. The strict limit with bounded responses gives $\partial_s\mathcal E_C=0$ and $\partial_s\mathcal P_A^C=0$. Spatial dynamics remains visible on the stretched interval for which $\tau=\epsilon s$ is held fixed: division of Eq.~\eqref{eq:wardE} and Eq.~\eqref{eq:wardP} by $\epsilon$ gives finite-clock conservation. A mode with finite-clock frequency $\Omega$ consequently has selected-clock frequency $\omega=\epsilon\Omega$. This is also the frequency window sampled by a thermal state with $T_C=\epsilon\widehat T_C$. The local equations and the probe calculation in Section~\ref{sec:scalar} therefore use the same order of time and frequency limits.

For a time-independent real function $f(x)$, the Brown--York charge associated with $f\xi_\lambda$ is \cite{Brown:1992br,Balasubramanian:1999re,deHaro:2000vlm,Skenderis:2002wp}
\begin{equation}
\begin{aligned}
 Q_\epsilon[f]&=\int_{\sphere^{d-1}}\dd^{d-1}x\sqrt h\,
 f\mathcal E_\epsilon,\\
 \frac{\dd Q_\epsilon[f]}{\dd s}
 &=-\epsilon\int_{\sphere^{d-1}}\dd^{d-1}x\sqrt h\,
 \mathcal P_\epsilon^A D_Af.
 \end{aligned}
 \label{eq:dQ}
\end{equation}
Its limit defines the smeared energy functional
\begin{equation}
 Q_C[f]=\int_{\sphere^{d-1}}\dd^{d-1}x\sqrt h\,
 f\mathcal E_C,
 \qquad Q_C[1]=E_C .
 \label{eq:Qf}
\end{equation}
For the Schwarzschild-AdS state, $\mathcal P_A^C=0$ and $\mathcal E_C$ is homogeneous. Hence $Q_C[f]=\bar f E_C$, where $\bar f=\Omega_{d-1}^{-1}\int\dd^{d-1}x\sqrt q\,f$. We use the sphere average itself as the constant harmonic coefficient; normalized spherical harmonics are defined separately in the scalar calculation. The homogeneous black-hole family consequently has a single nonzero angular energy component. At finite $\epsilon$, the global mode is conserved on a closed sphere, and the angular modes evolve with the flux specified in Eq.~\eqref{eq:dQ}.

The geometric action of $f(x)\partial_s$ preserves the static spatial metric and temporal vector, while the clock representative changes by $\mathcal L_{f\partial_s}\vartheta=\dd f$. Allowing this source transformation gives a covariant identity on the source family. For a functional of scalar insertions, denote by $\mathscr D_f^{\mathrm{src}}$ the variation of the sources, including the clock representative, and by $\mathscr D_f^{\rho}$ the variation of the state prescription. Diffeomorphism covariance gives
\begin{equation}
 \left[\sum_i f(x_i)\partial_{s_i}
       +\mathscr D_f^{\mathrm{src}}+\mathscr D_f^\rho\right]
 \left\langle\prod_i\Phi_i(s_i,x_i)\right\rangle=0.
 \label{eq:corrward}
\end{equation}
The source term is evaluated by functional differentiation of Eq.~\eqref{eq:sourcedVariation}, and includes the contact terms associated with the tensor weights of nonscalar insertions. In a stationary Gibbs state the constant time translation has vanishing source and state variations, giving dependence on time differences. An angularly dependent translation retains the source response in Eq.~\eqref{eq:corrward}. This is the sourced form relevant to the finite-frequency angular kernels below.

The distinction has a direct observable consequence. For a two-point function at separated angular points, stationarity gives $C=C(s_1-s_2;x_1,x_2)$. The insertion part of an angular translation acts as $[f(x_1)-f(x_2)]\partial_sC$. Its cancellation in Eq.~\eqref{eq:corrward} is supplied by the source and state responses. The calculated angular spectral kernel can therefore have time dependence at separated points while preserving covariance. For the constant mode, stationarity itself cancels the insertion terms, as required by energy conservation. This relation connects the smeared stress response to the frequency-diagonal thermal correlator and fixes which symmetry statement is used in the Mellin representation.

Equilibrium variations of the same global charge are determined by the extended Hamiltonian identity. With the generator held fixed and the source prescription inherited from Ref.~\cite{Xu:2026vpj}, the Brown--York and covariant phase-space charges agree at infinity. Using the finite coordinates of Eq.~\eqref{eq:finiteThermoCoordinates}, we obtain
\begin{equation}
\begin{aligned}
 \delta Q_C[1]&=\widehat T_C\delta S_C+V_C\delta P_C,\\
 \delta Q_C[f]&=\bar f\,\delta Q_C[1]
 \qquad\text{on the homogeneous family}.
 \end{aligned}
 \label{eq:firstlawward}
\end{equation}
The entropy contribution follows from the horizon Noether charge, and the pressure contribution comes from the cosmological-constant source \cite{Wald:1993nt,Iyer:1994ys,Iyer:1995kg,Kastor:2009wy,Dolan:2011xt,Xiao:2023lap}. Here $\delta$ changes the equilibrium solution at fixed contraction data. The derivative in Eq.~\eqref{eq:dQ} evolves a state in time. Both use $Q_C[1]$, so the thermodynamic and dynamical normalizations agree with the sourced stress response.

The comparison with null infinity is made at the level of source geometry and charge balance. Radiative Carrollian boundaries have their own energy flux through the boundary and corresponding BMS balance laws \cite{Bondi:1962px,Sachs:1962wk,Ashtekar:1981bq,Wald:1999wa,Barnich:2010eb,Barnich:2011mi,Strominger:2013jfa,He:2014laa,Donnay:2022aba,Donnay:2022wvx,Ruzziconi:2024kzo}. The thermal AdS cylinder used here has reflecting asymptotic boundary conditions and a horizon that absorbs the probe. Its stress conservation law is Eq.~\eqref{eq:wardE}, with the spatial flux inherited from the relativistic source family. This comparison identifies the common time-shift representation while retaining the physical boundary conditions that determine each response.

\section{Thermal scalar response and angular absorption}
\label{sec:scalar}

The stress and Gibbs constructions fix the clock, source normalization and thermal state. A scalar probe then measures the response to an inhomogeneous boundary perturbation through its ingoing horizon flux. We derive the spectral clock map before solving the radial problem, so the angular and frequency dependence can be separated from the large-rank normalization. The numerical example uses a massless scalar in four-dimensional Schwarzschild-AdS. Its transmission coefficients supply the spectral input for Section~\ref{sec:celestial}, while its source convention also applies to the type IIB operators in Section~\ref{sec:regge}. The spherical absorption problem and the planar four-point problem therefore test complementary observables within the same contraction prescription.

\subsection{Operator sources and the change of clock}
\label{subsec:celestial-modes}

Let $\Phi_\tau$ be the operator sourced by $j_\tau$ in the finite Lorentzian clock. The linear coupling transforms as
\begin{equation}
 \begin{gathered}
 \int\dd\tau\,\dd^{d-1}x\sqrt h\,j_\tau\Phi_\tau
 =\int\dd s\,\dd^{d-1}x\sqrt h\,j_s\Phi_s,\\
 j_s(s,x)=j_\tau(\epsilon s,x),\qquad
 \Phi_s(s,x)=\epsilon\Phi_\tau(\epsilon s,x).
 \end{gathered}
 \label{eq:scalarSourceMap}
\end{equation}
We use this source normalization for the probe. More generally, a prescribed operator map $\Phi_s=z_\epsilon\Phi_\tau(\epsilon s)$ requires $j_s=(\epsilon/z_\epsilon)j_\tau(\epsilon s)$. The operator factor and temporal Jacobian are therefore separate choices in the response functional. For a scalar spectral density defined as the Fourier transform of the commutator, their relation is
\begin{equation}
 \rho_{s,L}(\omega)
 =\frac{z_\epsilon^2}{\epsilon}\,
   \rho_{\tau,L}\left(\frac{\omega}{\epsilon}\right),
 \qquad
 \beta_s\omega=\beta_\tau\frac{\omega}{\epsilon}.
 \label{eq:spectralClockMap}
\end{equation}
This follows by the substitution $\tau=\epsilon s$ in the Fourier transform. The Bose occupation argument is preserved, as required by KMS. For the source choice in Eq.~\eqref{eq:scalarSourceMap}, $z_\epsilon=\epsilon$ and the prefactor in Eq.~\eqref{eq:spectralClockMap} is $\epsilon$.

For a Hermitian scalar, write $G^>_s(s;x,x')=\langle\Phi_s(s,x)\Phi_s(0,x')\rangle$ and use a Fourier transform over all real frequencies. Its positive-frequency part obeys $G^>_s(\omega)=[1+n_B(\beta_s\omega)]\rho_s(\omega)$, while $G^<_s(\omega)=n_B(\beta_s\omega)\rho_s(\omega)$ for $\omega>0$. Here $n_B(y)=(e^y-1)^{-1}$. The latter is the occupation contribution of the spectral density in the chosen thermal geometry. A comparison with a different vacuum geometry would also change that geometry's spectral function. This choice fixes the occupation moment used in Section~\ref{sec:celestial} and retains both correlators required by KMS.

\subsection{Horizon flux and angular transmission}
\label{subsec:probe-scalar-spectral-density}

Consider a minimally coupled probe with action
\begin{equation}
I_\varphi=-\frac{\mathcal N_\varphi(c)}{2}
 \int\dd^{d+1}x\sqrt{-g}\,
 \left[g^{\mu\nu}\partial_\mu\varphi\partial_\nu\varphi
       +m_\varphi^2\varphi^2\right].
 \label{eq:probeScalarAction}
\end{equation}
We choose $\mathcal N_\varphi(c)=\mathcal N_{\varphi C}/\epsilon$, where the positive constant $\mathcal N_{\varphi C}$ fixes the single-trace normalization. In the finite clock we decompose $\varphi=e^{-i\Omega\tau}Y_{LM}(x)\psi_L(r;\Omega)$, with orthonormal harmonics on the unit sphere. The radial equation is
\begin{equation}
 \frac{1}{r^{d-1}}\partial_r(r^{d-1}f\partial_r\psi_L)
 +\left[\frac{\Omega^2}{f}-\frac{L(L+d-2)}{r^2}
                   -m_\varphi^2\right]\psi_L=0.
 \label{eq:probeRadialEquation}
\end{equation}
The retarded prescription imposes an ingoing wave at the horizon and a prescribed nonnormalizable boundary coefficient \cite{Son:2002sd}. For a unit boundary source, the regular zero-frequency profile is denoted by $\psi_{0,L}$. Near the horizon the ingoing solution is proportional to $(r-r_h)^{-i\Omega/(4\pi T_\tau)}$. The radially conserved flux therefore determines the leading absorptive response:
\begin{equation}
\begin{aligned}
 \mathcal F_L&=\mathcal N_\varphi(c)r^{d-1}f\,
       \operatorname{Im}(\psi_L^*\partial_r\psi_L),\\
 \rho_{\tau,L}(\Omega)&=-2\operatorname{Im}G^R_{\tau,L}(\Omega)
 =\frac{A_{L,C}}{\epsilon}\Omega+O(\epsilon^{-1}\Omega^3),\\
 A_{L,C}&=2\mathcal N_{\varphi C}r_h^{d-1}|\psi_{0,L}(r_h)|^2.
 \end{aligned}
 \label{eq:rhoLLowOmega}
\end{equation}
The odd-frequency expansion assumes a smooth nonextremal channel around zero frequency. Positive $\Omega$ gives positive $\rho_{\tau,L}$ for the scalar norm in Eq.~\eqref{eq:probeScalarAction}. For $m_\varphi=L=0$, horizon regularity makes $\psi_{0,0}=1$, and the leading coefficient is governed by the horizon area. The horizon-area dependence agrees with the low-energy absorption universality in asymptotically flat black-hole backgrounds \cite{Das:1996we}; Eq.~\eqref{eq:rhoLLowOmega} fixes the AdS source normalization through the radial flux. The factors for $L>0$ measure the angular transmission from the boundary to the horizon and carry additional dependence on the state.

For the four-dimensional massless problem, let $x=r_h/\ell$ and define $\mathcal R_L(x)=|\psi_{0,L}(r_h)|^2$. A horizon-normalized solution $y_L(r_h)=1$ obeys
\begin{equation}
\begin{gathered}
 (r^2f y_L')'=L(L+1)y_L,\\[3pt]
 y_L'(r_h)=\frac{L(L+1)}{r_h^2f'(r_h)},
 \qquad \mathcal R_L=|y_L(\infty)|^{-2}.
 \end{gathered}
 \label{eq:zeroModeTransmission}
\end{equation}
For $L>0$, the positive horizon value and the integrated radial equation imply $y'_L>0$ throughout the exterior. Thus the unit-source horizon amplitude lies between zero and one. The suppression has a direct physical interpretation: angular gradients reduce the field that reaches the absorbing surface for a fixed boundary source. The homogeneous mode retains unit transmission.

The large-radius behavior can be derived by writing $r=r_h z$ and expanding at fixed $L$. The leading integrated radial equation gives
\begin{equation}
\begin{aligned}
 y_L(\infty)&=1+\frac{L(L+1)}{x^2}\mathcal J+O(x^{-4}),\\
 \mathcal J&=\int_1^\infty\frac{\dd z}{z(z^2+z+1)}
 =\frac12\log3-\frac{\pi}{6\sqrt3}.
 \end{aligned}
 \label{eq:largeRadiusTransmission}
\end{equation}
and therefore $\mathcal R_L=1-2L(L+1)\mathcal J/x^2+O(x^{-4})$. The expansion is organized by $L(L+1)/x^2$. It quantifies the approach to the homogeneous response for a fixed angular harmonic on a large black hole. The angular structure thus becomes progressively less suppressed as the horizon radius grows compared with the boundary curvature scale.

\subsection{Numerical scalar response}
\label{subsec:scalar-numerics}

We solve Eq.~\eqref{eq:zeroModeTransmission} on the stable spherical branch and normalize the solution by its asymptotic source coefficient. For the massless field the boundary expansion is $y_L=A_L[1-L(L+1)\ell^2/(2r^2)]+B_Lr^{-3}+O(r^{-4})$. The results in Table~\ref{tab:scalarTransmission} quantify the angular dependence over $0.6\leq x\leq4$. Near $x=0.6$, the dipole coefficient is $0.1924$ and the quadrupole coefficient is $0.01535$. At $x=4$ they rise to $0.9411$ and $0.8349$. The higher harmonics are consequently much more sensitive to the horizon radius than the homogeneous mode. At the Hawking--Page radius, the values $0.4549$ and $0.1180$ provide distinct response coefficients for angular sources in the same thermodynamic state.

\begin{table}[tb]
 \centering
 \begin{tabular}{ccccc}
 \toprule
 $x=r_h/\ell$ & $\mathcal R_0$ & $\mathcal R_1$ & $\mathcal R_2$ & $\mathcal R_3$\\
 \midrule
 $0.6$ & $1$ & $0.1924$ & $0.01535$ & $8.168\times10^{-4}$\\
 $1$   & $1$ & $0.4549$ & $0.1180$  & $0.02185$\\
 $2$   & $1$ & $0.7931$ & $0.5112$  & $0.2779$\\
 $4$   & $1$ & $0.9411$ & $0.8349$  & $0.7007$\\
 \bottomrule
 \end{tabular}
 \caption{Zero-frequency angular transmission for a massless probe in four-dimensional Schwarzschild-AdS. The coefficients multiply the homogeneous horizon-area factor in Eq.~\eqref{eq:rhoLLowOmega}.}
 \label{tab:scalarTransmission}
\end{table}

The finite-frequency calculation uses $\psi_L=e^{-i\Omega r_*}y_L$ to factor out the ingoing phase. Here $\dd r_*/\dd r=f^{-1}$ and $r_*(\infty)=0$. The regular radial equation and horizon derivative are
\begin{equation}
\begin{gathered}
 r^2f y_L''+(2rf+r^2f'-2i\Omega r^2)y_L'
       -[2i\Omega r+L(L+1)]y_L=0,\\[4pt]
 y_L'(r_h)=\frac{2i\Omega r_h+L(L+1)}
 {r_h^2[f'(r_h)-2i\Omega]}y_L(r_h).
 \end{gathered}
 \label{eq:regularIngoingEquation}
\end{equation}
With $y_L(r_h)=1$, its boundary source coefficient $A_L(\Omega)$ gives
\begin{equation}
 \mathcal R_L(x,\Omega)
 :=\frac{\rho_{\tau,L}(\Omega)}
 {2\mathcal N_\varphi(c)r_h^2\Omega}
 =|A_L(\Omega)|^{-2}.
 \label{eq:finiteFrequencyTransmission}
\end{equation}
The radial flux remains constant throughout the exterior to the reported accuracy. Increasing the extraction radius and tightening the radial tolerance changes the tabulated coefficients by less than $10^{-8}$; the normalized flux variation is below $2\times10^{-8}$ at the representative frequency points. We report four significant digits, which resolve the physical differences between the angular channels.

\begin{figure}[tb]
 \centering
 \begin{subfigure}{0.49\linewidth}
 \centering\includegraphics[width=\linewidth]{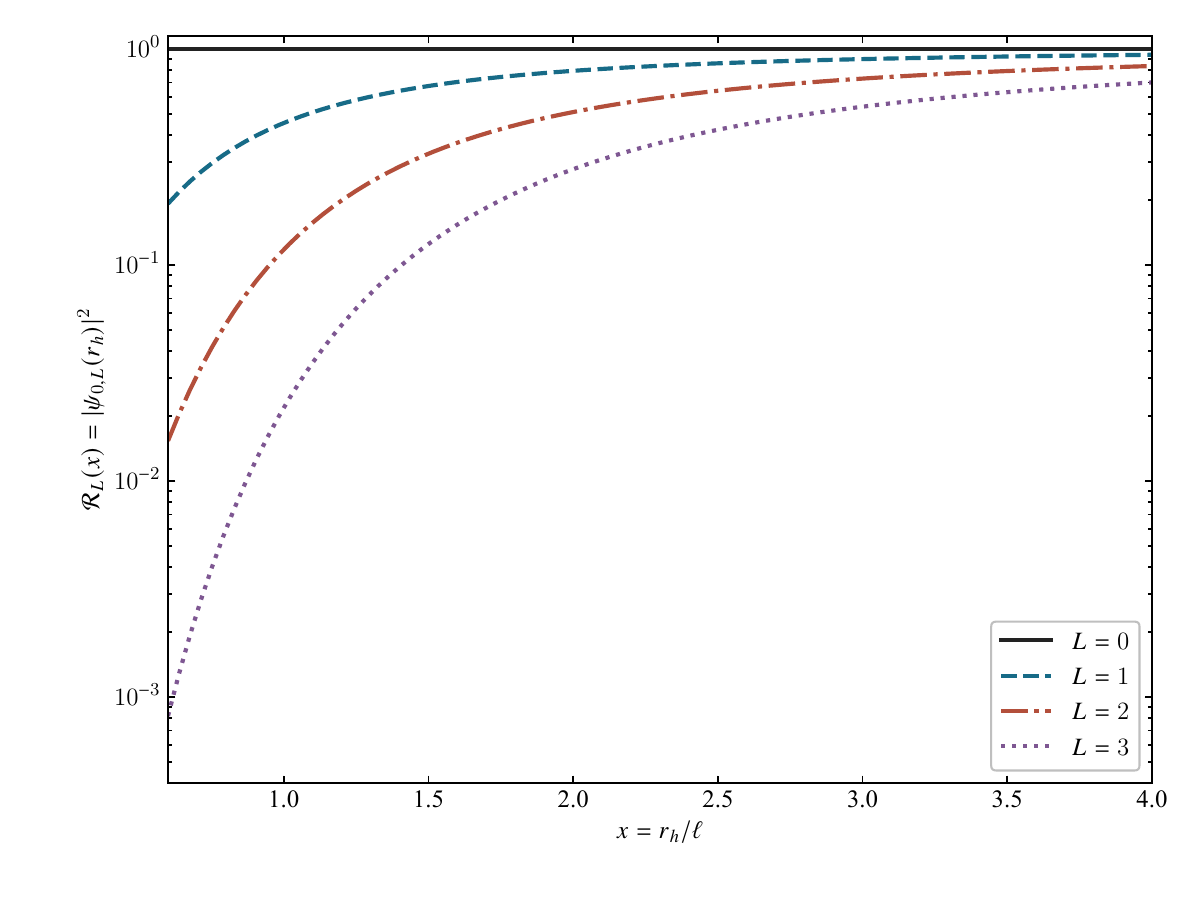}
 \caption{Angular transmission at zero frequency.}
 \end{subfigure}\hfill
 \begin{subfigure}{0.49\linewidth}
 \centering\includegraphics[width=\linewidth]{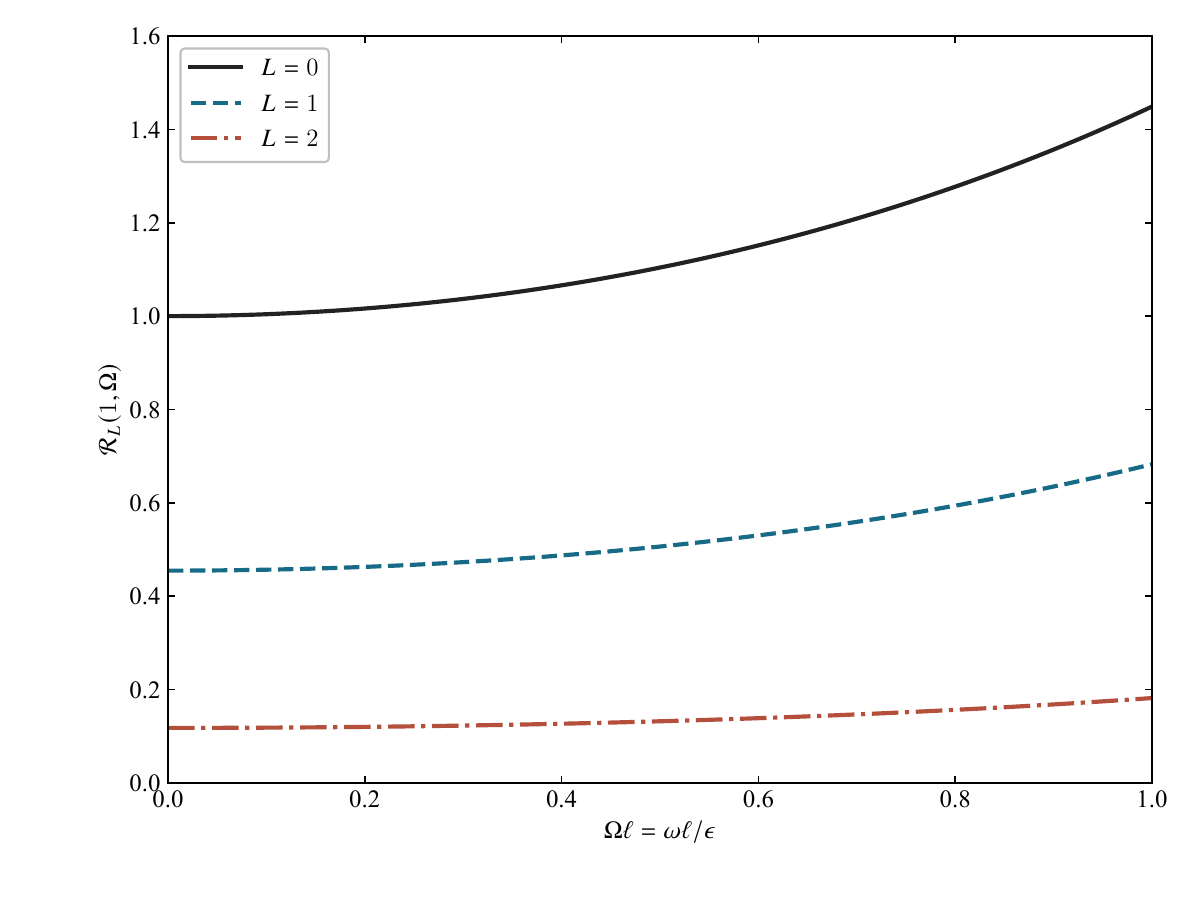}
 \caption{Frequency dependence at $r_h=\ell$.}
 \end{subfigure}
 \caption{Massless scalar response of the four-dimensional Schwarzschild-AdS family. The left panel shows the approach of fixed angular harmonics to the homogeneous response as the horizon grows. The right panel shows the spectral response normalized by the horizon-area slope, with $\Omega=\omega/\epsilon$. The same curves therefore determine the thermal frequency window of the contracted ensemble.}
 \label{fig:scalarResponse}
\end{figure}

Figure~\ref{fig:scalarResponse} displays both the zero-frequency angular filtering and the finite-frequency response at $x=1$. At $\Omega\ell=0.2$, the values for $L=0,1,2$ are $1.016$, $0.4629$ and $0.1202$. At $\Omega\ell=1$, they become $1.449$, $0.6828$ and $0.1820$. The response therefore increases with frequency over the displayed interval while retaining the angular hierarchy. The departure from the linear spectral approximation at $\Omega\ell=0.2$ is below two per cent in these channels. At the larger frequency, the full radial solution is needed to determine the thermal moment. These values identify the physical size of the low-frequency interval used for the moments in Section~\ref{sec:celestial}.

\section{A thermal four-point limit with finite string spreading}
\label{sec:regge}

The source map now permits a normalized interacting observable to be compared along the same large-rank family. Its relevant scale is the time at which the large-rank suppression of a connected four-point function is compensated by the horizon boost. We examine this scale in the type IIB family of Section~\ref{subsec:AdS5CFT4}. The calculation retains the AdS radius and finite-clock temperature and uses the planar black-brane state. Its blackening function is $f(r)=r^2\ell^{-2}(1-r_h^4/r^4)$, its temperature is $\widehat T_C=r_h/(\pi\ell^2)$, and its boundary spatial metric is flat. The rank, coupling and observation time can therefore be varied independently in a thermal field-theory setting before selecting their correlated limit. The finite thermodynamic energy fixes the rank normalization, and the four-point function determines which coupling trajectories preserve its gravitational interaction.

\subsection{The observable and the logarithmic time window}

Choose scalar single-trace operators $V$ and $W$ sourced by the $S^5$-singlet dilaton and axion. Both have conformal dimension four in the parent super Yang--Mills theory, with their normalization defined by the bulk sources \cite{Maldacena:1997re,Gubser:1998bc,Witten:1998qj}. The constant internal harmonic therefore selects the singlet Regge exchange. We subtract their thermal one-point functions. The thermal period is $\beta_s=1/T_C=1/(\epsilon\widehat T_C)$. For distinct imaginary-time regulators at fixed fractions of this period, define the normalized out-of-time-order correlator
\begin{equation}
 \mathcal F_\epsilon(s,\mathbf b_0)
 =\frac{\langle V_{s,1}W_{s,2}V_{s,3}W_{s,4}\rangle_{\beta_s}}
 {\langle V_{s,1}V_{s,3}\rangle_{\beta_s}
  \langle W_{s,2}W_{s,4}\rangle_{\beta_s}}.
 \label{eq:regulatedFourPoint}
\end{equation}
The real times of insertions $1,3$ are zero and those of $2,4$ are $s$; the two pairs have transverse separation $\mathbf b_0$ in the thermal coordinates defined below. Fixed smearing profiles keep the insertion energies of thermal order. The source factors in Eq.~\eqref{eq:scalarSourceMap} cancel between the numerator and denominator, so this observable has the same normalization in the two clocks. Its finite-clock time argument is $\tau=\epsilon s$, and its disconnected large-rank value is one.

Let $\lambda$ denote the 't Hooft coupling and introduce
\begin{equation}
\begin{gathered}
 \mathcal L_\epsilon=\log\epsilon^{-1},\qquad
 u=2\pi\widehat T_C\epsilon s-\mathcal L_\epsilon,\\[3pt]
 \frac{\alpha'}{\ell^2}=\lambda^{-1/2},\qquad
 g_s=\frac{\lambda\sqrt\epsilon}{4\pi N_C}.
 \end{gathered}
 \label{eq:reggeScalingVariables}
\end{equation}
The rank relation is Eq.~\eqref{eq:AdS5N2}. At fixed $u$, the selected time grows as $\epsilon^{-1}\mathcal L_\epsilon$, exceeding the thermal interval by a logarithmic factor. Constants such as the entropy per thermal cell shift $u$ by a finite amount. At a fixed thermal time the connected correlator is of order $\epsilon$; in this logarithmic window its leading eikonal phase can be finite.

We use the leading curved-horizon Regge amplitude of Shenker and Stanford \cite{Shenker:2014cwa}. Their result resums the enhancement of the string correction by the relative horizon boost. In coordinates $\mathbf X=r_h\mathbf x_\partial/\ell^2$, let $b=|\mathbf X-\mathbf Y|$ and $m_\perp^2=6$. If $\mathfrak s$ is the local squared center-of-mass energy in units of $\ell^{-2}$, their amplitude becomes
\begin{equation}
\begin{aligned}
 \delta_\epsilon(\mathfrak s,b)
 &\simeq\frac{2\pi^2\epsilon}{N_C^2}\,\mathfrak s\,
 \mathcal K_{\zeta_\epsilon}(b),\\
 \mathcal K_\zeta(b)
 &=\int\frac{\dd^3\mathbf k}{(2\pi)^3}
 \frac{e^{i\mathbf k\cdot\mathbf b-\zeta(\mathbf k^2+m_\perp^2)}}
 {\mathbf k^2+m_\perp^2},\\
 \zeta_\epsilon&=
 \frac{\log[\mathfrak s/(4\sqrt\lambda)]-i\pi/2}{2\sqrt\lambda}.
 \end{aligned}
 \label{eq:parentReggeKernel}
\end{equation}
The coefficient follows from $G_5/\ell^3=\pi/(2N^2)$ and the transverse coordinate change. The mass $m_\perp$ is the horizon screening scale in these units. The Regge approximation retains the graviton pole and the leading string correction to its trajectory; the corresponding elastic amplitude enters the correlator through its horizon wave functions.

Write $\mathfrak s=pq\exp(\mathcal L_\epsilon+u)$, where $p,q>0$ are dimensionless unboosted null momenta. At fixed $p,q,u$, the strength $\epsilon\mathfrak s=pq e^u$ is finite, while the accumulated string correction depends on $\mathcal L_\epsilon/\sqrt\lambda$. Consequently, a sufficient window for the Einstein four-point limit and weak string coupling is
\begin{equation}
 \mathcal L_\epsilon^2\ll\lambda\ll\epsilon^{-1/2}.
 \label{eq:EinsteinScramblingWindow}
\end{equation}
The lower bound comes from the boosted observable; the upper bound comes from $g_s\to0$. Equilibrium at a fixed large coupling can have a small curvature correction throughout the rank limit, while its four-point function accumulates that correction over the logarithmic time interval. The logarithmic observable therefore supplies an additional coupling condition beyond equilibrium control.

A distinct family lies at the lower edge of Eq.~\eqref{eq:EinsteinScramblingWindow}:
\begin{equation}
\begin{aligned}
 \lambda&=\kappa\mathcal L_\epsilon^2,\qquad 0<\kappa<\infty,\\
 \zeta_\epsilon&\longrightarrow\zeta_\kappa=\frac{1}{2\sqrt\kappa},\\
 \delta_\kappa(u,pq,b)
 &=\frac{2\pi^2}{N_C^2}pq e^u\mathcal K_{\zeta_\kappa}(b).
 \end{aligned}
 \label{eq:marginalReggeLimit}
\end{equation}
Here $g_s$ tends to zero, $\alpha'/\ell^2$ tends to zero, and the transverse string profile remains finite. The subleading term in the Regge trajectory contributes at order $\mathcal L_\epsilon/\lambda$, which vanishes on this family. The logarithm of $4\sqrt\lambda$ and the unboosted momenta give corrections of order $\log\mathcal L_\epsilon/\mathcal L_\epsilon$ for bounded nonzero momenta. The Regge signature phase also vanishes. These estimates identify the surviving term in the correlated limit. Taking the string length to zero at a fixed observation time yields the Einstein profile; allowing the observation time to grow with the rank retains the finite correction in Eq.~\eqref{eq:marginalReggeLimit}.

Let $w_V(p,\mathbf X)$ and $w_W(q,\mathbf Y-\mathbf b_0)$ denote the normalized horizon overlap weights of the chosen regulated insertions, including the null momentum measure. They are obtained from the finite-clock bulk-to-boundary wave functions, and their phases retain the insertion ordering. For each pair the integrated weight is one. The eikonal representation then gives the limiting boundary functional
\begin{equation}
 \begin{aligned}
 \mathcal F_\kappa(u,\mathbf b_0)
 &=\int_0^\infty\dd p\,\dd q\int\dd^3\mathbf X\,\dd^3\mathbf Y\,
 w_V(p,\mathbf X)w_W(q,\mathbf Y-\mathbf b_0)\\
 &\quad\times\exp\!\left[
 \frac{2\pi^2 i}{N_C^2}pq e^u
 \mathcal K_{\zeta_\kappa}(|\mathbf X-\mathbf Y|)\right].
 \end{aligned}
 \label{eq:limitingFourPointFunctional}
\end{equation}
Smooth regulated wave packets give integrable overlap weights, so bounded momentum intervals determine the limit and their tails can be taken afterwards. The weights approach their supergravity values as $\lambda$ grows. Inelastic corrections to the elastic overlap are suppressed in the regime with finite gravitational scattering strength \cite{Shenker:2014cwa}. Eq.~\eqref{eq:limitingFourPointFunctional} specifies the interaction kernel for the dilaton and axion correlator; the external wave functions encode the chosen boundary insertions. It also shows how the clock factors cancel for a normalized interacting observable, while the logarithmic observation time compensates the rank suppression.

\subsection{The surviving transverse profile}

For positive real $\zeta$, the denominator in Eq.~\eqref{eq:parentReggeKernel} can be integrated with a Schwinger parameter. Performing the transverse Gaussian integral yields
\begin{equation}
\begin{aligned}
 \mathcal K_\zeta(b)
 &=\int_\zeta^\infty\frac{\dd t}{(4\pi t)^{3/2}}
       \exp\!\left[-m_\perp^2t-\frac{b^2}{4t}\right]\\
 &=\frac{e^{-m_\perp b}}{8\pi b}\operatorname{erfc}\!\left(
 m_\perp\sqrt\zeta-\frac{b}{2\sqrt\zeta}\right)\\
 &\quad-\frac{e^{m_\perp b}}{8\pi b}\operatorname{erfc}\!\left(
 m_\perp\sqrt\zeta+\frac{b}{2\sqrt\zeta}\right).
 \end{aligned}
 \label{eq:finiteReggeProfile}
\end{equation}
The integral defines the continuous value at $b=0$. At $\zeta=0$ and $b>0$ it gives the Einstein profile $\mathcal K_0(b)=e^{-m_\perp b}/(4\pi b)$. For every $\zeta>0$ the center is finite. Differentiating the integral gives
\begin{equation}
 (-\nabla_{\mathbf b}^2+m_\perp^2)\mathcal K_\zeta(b)
 =\frac{e^{-m_\perp^2\zeta-b^2/(4\zeta)}}{(4\pi\zeta)^{3/2}}
 =-\partial_\zeta\mathcal K_\zeta(b).
 \label{eq:ReggeSmearedSource}
\end{equation}
Thus the finite interaction solves the screened horizon equation with a Gaussian source whose total strength decreases with $\zeta$. Eq.~\eqref{eq:finiteReggeProfile} retains both this central region and the screened tail. For $b$ large compared with $m_\perp\zeta$, the ratio $\mathcal K_\zeta(b)/\mathcal K_0(b)$ approaches one. The attenuation is concentrated in the central region, and the screened tail retains the Einstein normalization.

The zero-momentum kernel and its curvature give two integrated observables:
\begin{equation}
 \begin{aligned}
 \mathcal A_\kappa
 &\equiv m_\perp^2\int\dd^3\mathbf b\,
                      \mathcal K_{\zeta_\kappa}(b)
   =e^{-3/\sqrt\kappa},\\
 \langle b^2\rangle_\kappa
 &\equiv\frac{\int\dd^3\mathbf b\,b^2\mathcal K_{\zeta_\kappa}(b)}
              {\int\dd^3\mathbf b\,\mathcal K_{\zeta_\kappa}(b)}
   =1+\frac{3}{\sqrt\kappa}.
 \end{aligned}
 \label{eq:ReggeProfileMoments}
\end{equation}
Applying $-\nabla_{\mathbf k}^2$ to the Fourier kernel at zero momentum gives the mean square radius. In boundary length units this radius is $\langle b^2\rangle_\kappa/(\pi\widehat T_C)^2$. The attenuation and width satisfy \mbox{$-\log\mathcal A_\kappa=\langle b^2\rangle_\kappa-1$}. They therefore measure the same accumulated Regge correction through two integrated properties. These are moments of the interaction kernel, with the physical correlator obtained by the convolution in Eq.~\eqref{eq:limitingFourPointFunctional}.

\begin{figure}[t]
 \centering
 \begin{subfigure}{0.49\linewidth}
  \includegraphics[width=\linewidth]{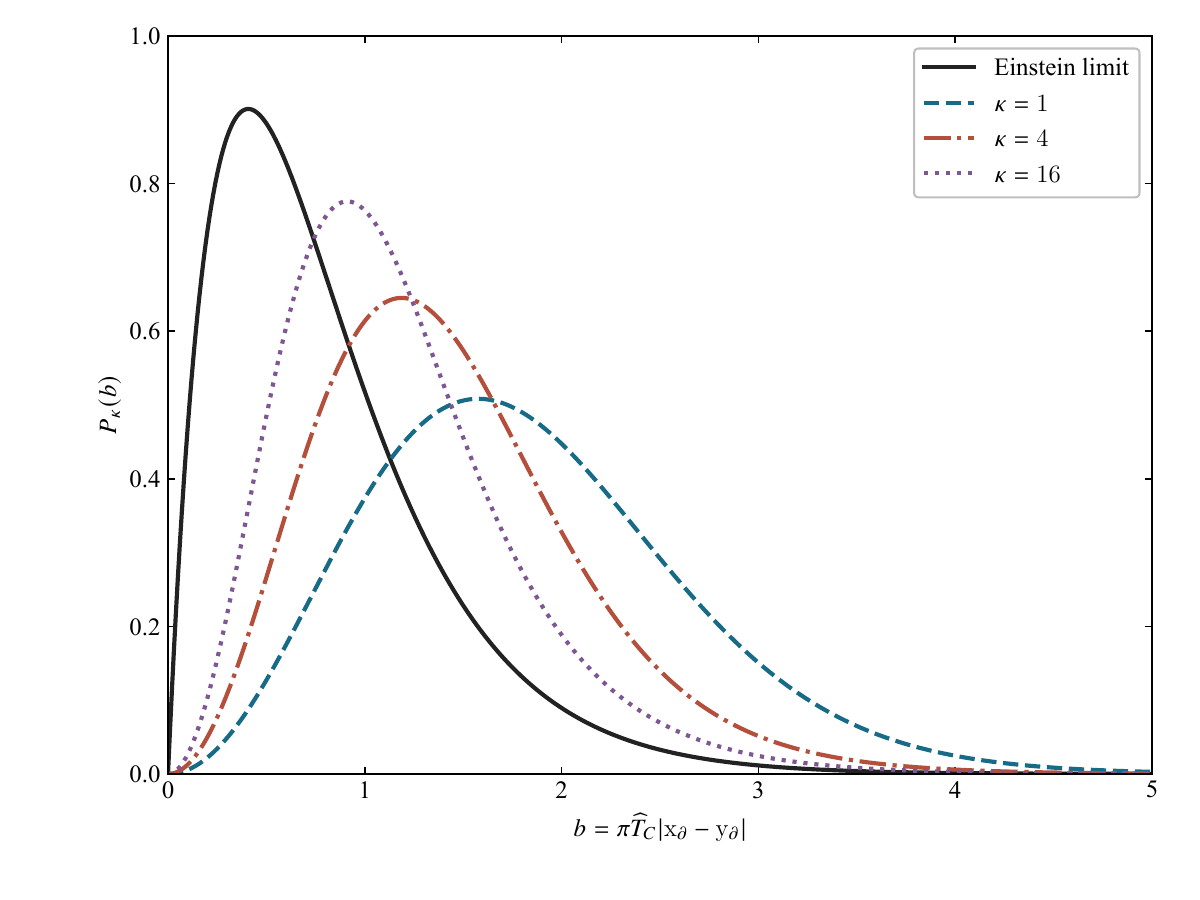}
  \caption{Normalized radial interaction profile.}
 \end{subfigure}\hfill
 \begin{subfigure}{0.49\linewidth}
  \includegraphics[width=\linewidth]{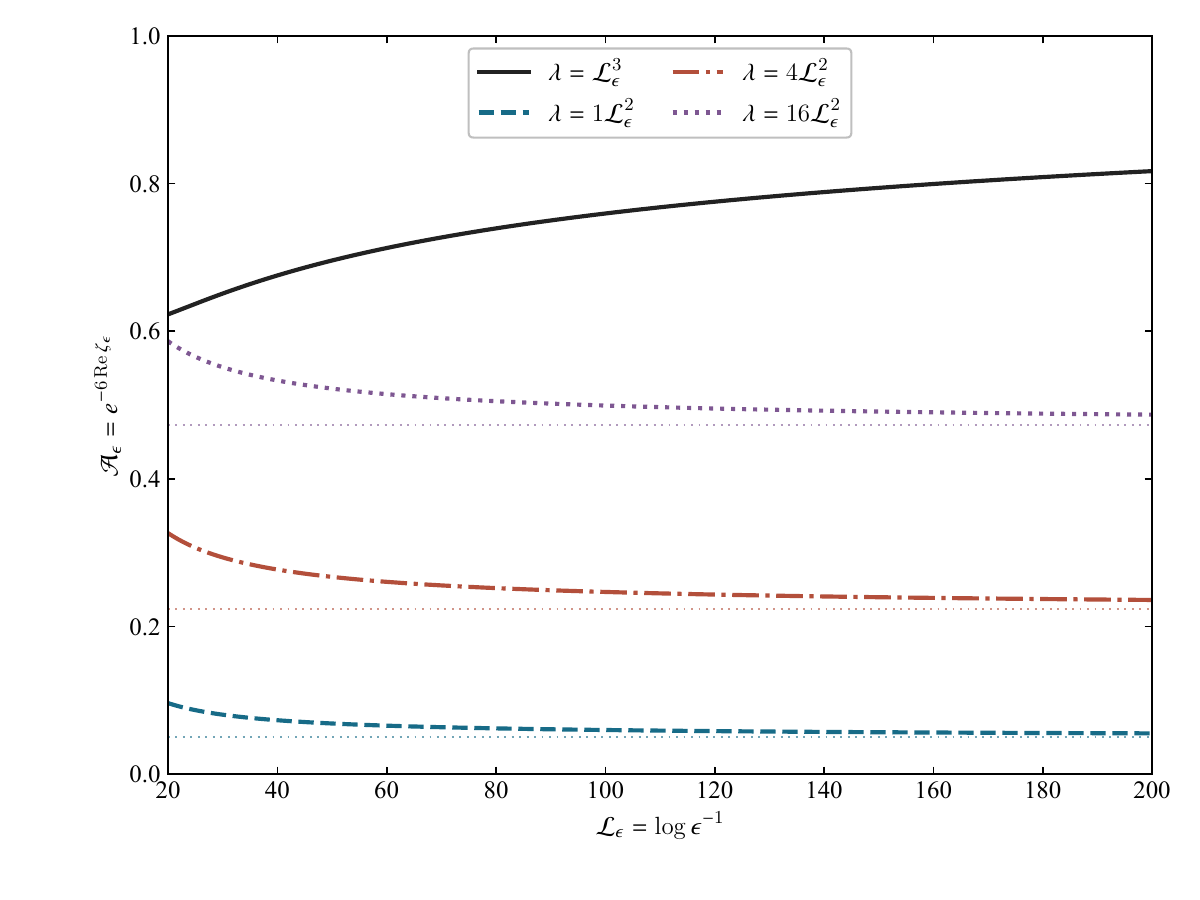}
  \caption{Integrated strength along the rank limit.}
 \end{subfigure}
 \caption{The four-point interaction on the logarithmic Carrollian time interval. The radial profile is $P_\kappa(b)=4\pi b^2m_\perp^2e^{m_\perp^2\zeta_\kappa}\mathcal K_{\zeta_\kappa}(b)$ and has unit integral. Its finite width records string spreading as $\alpha'/\ell^2\to0$. The right panel shows $\mathcal A_\epsilon=\exp[-6\operatorname{Re}\zeta_\epsilon]$ at $pq=1$ and $u=0$, including the string-energy logarithm. Dotted lines mark the finite-$\kappa$ limits. The family $\lambda=\mathcal L_\epsilon^3$ lies in the Einstein window and approaches unit strength.}
 \label{fig:reggeLimit}
\end{figure}

Figure~\ref{fig:reggeLimit} displays the radial distribution of this kernel and the approach to its finite strength. For $\kappa=1,4,16$, the strengths are $0.0498$, $0.2231$ and $0.4724$, while the root mean square radii are $2.000$, $1.581$ and $1.323$. The Einstein values are one for both quantities. The finite-parameter curves retain the string-energy logarithm in Eq.~\eqref{eq:parentReggeKernel}; their convergence is logarithmic, even though the string coupling decreases exponentially with $\mathcal L_\epsilon$. Evaluation of the Schwinger integral agrees with the complementary-error-function form, and integrating the radial profiles yields the moments in Eq.~\eqref{eq:ReggeProfileMoments}.

The logarithmic window also separates growth rates from accumulated effects. The leading zero-transverse-momentum Regge growth rate in the selected clock is
\begin{equation}
 \frac{\lambda_{R,s}}{2\pi T_C}
 =1-\frac{3}{\sqrt\lambda}+O(\lambda^{-1}).
 \label{eq:ReggeGrowthRate}
\end{equation}
This ratio tends to one in both Eq.~\eqref{eq:EinsteinScramblingWindow} and Eq.~\eqref{eq:marginalReggeLimit}. Their four-point kernels differ by a finite spatial profile. At zero transverse momentum the marginal family shifts the logarithmic onset variable by $3/\sqrt\kappa$. At fixed impact parameter the shift is determined by the full kernel and the external wave functions. Thus the leading growth rate and equilibrium thermodynamics leave room for a continuous family of distinct normalized four-point limits.

Recent constructions clarify the holographic setting of this result. Bagchi and collaborators use the Okuda--Penedones large-radius limit of vacuum string amplitudes to define Carrollian correlators at null infinity \cite{Bagchi:2026doublescaled}. Their kinematic scaling retains flat-space scattering data. Here the AdS radius and thermal state remain fixed, and the accumulated horizon boost retains a thermal Regge interaction. Fontanella and Payne develop a contraction of the microscopic string and gauge descriptions \cite{Fontanella:2025tbs}; our observable is defined along the relativistic type IIB sequence with the specified rank and clock normalization. Gao and Liu organize pole-skipping data into Regge trajectories in Rindler conformal theories and the Sachdev--Ye--Kitaev chain \cite{Gao:2025stringy}. Eq.~\eqref{eq:ReggeGrowthRate} identifies the trajectory information entering our thermal amplitude, while Eq.~\eqref{eq:finiteReggeProfile} gives the finite spatial kernel that survives its accumulation over the rank-dependent time interval. The resulting functional ties the thermodynamic rank normalization to an interacting holographic observable. Its finite spatial profile records the coupling trajectory after the leading thermal state and local growth rate have reached their common limits.

\section{Fourier--Mellin observables and time-shift identities}
\label{sec:celestial}

The preceding sections determine thermal response in frequency space and an interacting observable on the scrambling interval. We now express the stationary spectral data in a basis adapted to time translations. The Fourier--Mellin map acts on the computed scalar kernel, retains its finite thermal window and converts frequency multiplication into a shift of weight. Its application to a supplied three-point channel fixes the corresponding normalization and independent frequency measures.

Time shifts connect the thermal frequency representation to the source geometry of celestial constructions. The BMS and null-infinity literature establishes the geometric setting for angle-dependent translations and radiative charges \cite{Bondi:1962px,Sachs:1962wk,Sachs:1962,Newman:1961qr,Newman:1962cia,Penrose:1962ij,Penrose:1965am,Geroch:1977big,Ashtekar:1981bq,Ashtekar:1987tt,Ashtekar:1981sf,Madler:2016xju,Winicour:2008vpn,Wald:1999wa,Ashtekar:2001is,Barnich:2001jy,Barnich:2010eb,Barnich:2011mi,Barnich:2010ojg,Campiglia:2015yka,Campiglia:2016efb,Compere:2018ylh,Compere:2020lrt,Chandrasekaran:2018aop,Prabhu:2019fsp,Ashtekar:2024stm,Harlow:2019yfa}. In scattering, soft theorems and memory relate the corresponding charges to infrared observables \cite{Weinberg:1965nx,Low:1958,Burnett:1967km,Gross:1968in,Strominger:2013jfa,Strominger:2014pwa,Strominger:2013lka,He:2014laa,Lysov:2014csa,Cachazo:2014fwa,Campiglia:2014yka,Kapec:2014opa,He:2017fsb,Donnay:2018neh,Strominger:2017zoo,Nguyen:2021ydb,Arkani-Hamed:2020gyp,Kapec:2016jld,Fiorucci:2023lpb,Banerjee:2022wht,Nguyen:2025zhg,Kulkarni:2025qcx,Ruzziconi:2026bix}. These relations motivate the Mellin basis used in celestial holography \cite{Pasterski:2021rjz,Raclariu:2021zjz,Pasterski:2021raf}. For the thermal AdS boundary, the time-shift identity is the sourced relation in Eq.~\eqref{eq:corrward}. Its angular dependence includes the response of the boundary metric and the state, and this dependence must accompany the frequency transform.

Carrollian correlators have been related to scattering amplitudes and sourced celestial Ward identities \cite{Bagchi:2022emh,Donnay:2022aba,Donnay:2022wvx,Bagchi:2023cen,Bagchi:2024gnn,Ruzziconi:2024kzo,Poulias:2025eck}. Related developments concern conformal Carroll dynamics, null-boundary theories and higher-dimensional symmetry structures \cite{Bagchi:2019xfx,Bagchi:2019clu,Bagchi:2024qsb,Aggarwal:2025hji,PhysRevLett.105.171601,Bagchi:2012cy,Bagchi:2021gai,Ciambelli:2018wre,Campoleoni:2018ltl,Bagchi:2016bcd,Dutta:2022vkg,Have:2024dff,Saha:2023abr,Geiller:2024bgf,Freidel:2021fxf,Sudhakar:2023uan,Fiorucci:2025twa,Himwich:2019dug,Guevara:2021abz,Jiang:2021csc}. Correlator-based celestial prescriptions and finite-temperature examples further develop this representation \cite{Pasterski:2022chaos,Sleight:2024revisited,Iacobacci:2024KL,Long:2025thermal}.

We keep the dimensional generator clock $s$ for the correlators. Rescaling the boundary spatial metric to the unit-sphere representative changes the angular measure, while $Q_C[f]=\int\dd^{d-1}x\sqrt q\,f\varepsilon_C$ with $\varepsilon_C=\ell^{d-1}\mathcal E_C$. This convention leaves the generator and inverse temperature in the units of Eq.~\eqref{eq:xilambdaG}. The Mellin weight labels the frequency transform. In scattering applications, an additional angular conformal transformation law identifies these weights with celestial conformal dimensions \cite{Pasterski:2021rjz,Raclariu:2021zjz,Pasterski:2021raf,Bagchi:2022emh,Donnay:2022aba,Donnay:2022wvx}. Here the thermal sphere and black-hole state determine the angular kernel. The transform is useful independently of that additional representation-theory identification and can be compared with correlator-based celestial prescriptions \cite{Bagchi:2023cen,Bagchi:2024gnn,Ruzziconi:2024kzo,Poulias:2025eck,Pasterski:2022chaos,Sleight:2024revisited,Iacobacci:2024KL,Long:2025thermal}.

\subsection{Frequency modes and the thermal pairing}
\label{subsec:mellin-definitions}

We define frequency modes and their Mellin transforms by
\begin{equation}
\begin{aligned}
 \Phi^\eta_s(\omega,x)&=\int_{-\infty}^{\infty}\dd s\,
 e^{i\eta\omega s}\Phi_s(s,x),\\
 \mathcal O^\eta_\Delta(x)&=\int_0^\infty\dd\omega\,
 \omega^{\Delta-1}\Phi^\eta_s(\omega,x),\qquad \eta=\pm1.
 \end{aligned}
 \label{eq:celestialoperator}
\end{equation}
Stationarity fixes the pairing between positive and negative frequencies:
\begin{equation}
 \langle\Phi_s^+(\omega,x)\Phi_s^-(\omega',x')\rangle
 =2\pi\delta(\omega-\omega')G^>_s(\omega;x,x').
 \label{eq:frequencyKernelPlusMinus}
\end{equation}
Accordingly, the Mellin two-point function contains $2\pi\int_0^\infty\dd\omega\,\omega^{\Delta+\Delta'-2}G^>_s(\omega)$. The same factor multiplies its occupation contribution. Two modes with the same frequency sign instead produce a delta function supported at the sum of their frequencies. For positive frequencies this identifies the conjugate pairing as the appropriate nonzero thermal kernel.

\subsection{Finite thermal Mellin moments}
\label{subsec:probe-mellin}

The source map in Eq.~\eqref{eq:scalarSourceMap} and the horizon coefficient in Eq.~\eqref{eq:rhoLLowOmega} give
\begin{equation}
 \rho_{s,L}(\omega)
 =\frac{A_{L,C}}{\epsilon}\omega
 +O\!\left(\frac{\omega^3}{\epsilon^3}\right),
 \qquad \omega=\epsilon\Omega .
 \label{eq:probeThermalKernel}
\end{equation}
The remainder is ordered at fixed small $\Omega$; the interval of validity in $\omega$ shrinks with $\epsilon$. Let $\Lambda_s=\epsilon\widehat\Lambda$ and $\beta_s=\widehat\beta_C/\epsilon$. For $a=\Delta+\Delta'$ with $\operatorname{Re}a>1$, define the finite-window occupation moment by
\begin{equation}
 \mathcal G^{\mathrm{occ,IR}}_{\Delta,\Delta';LM}
 =2\pi\int_0^{\Lambda_s}\dd\omega\,
 \frac{\omega^{a-2}\rho_{s,L}(\omega)}{e^{\beta_s\omega}-1}.
 \label{eq:probeMellinBeforeScaling}
\end{equation}
Orthonormality gives the factor $\delta_{LL'}\delta_{MM'}$ for a mode paired with its complex conjugate. We display the diagonal coefficient. Substituting the leading spectral term produces
\begin{equation}
 \begin{aligned}
 \epsilon^{1-a}\mathcal G^{\mathrm{occ,IR}}_{\Delta,\Delta';LM}
 &=2\pi A_{L,C}\widehat\beta_C^{-a}
       M(a;\widehat\beta_C\widehat\Lambda)
       +\hbox{higher-frequency contributions},\\
 M(a;X)&=\int_0^X\dd y\,\frac{y^{a-1}}{e^y-1}.
 \end{aligned}
 \label{eq:probeDoubleScaledHat}
\end{equation}
The contraction power arises from the integration measure, the two Mellin weights, the operator normalization and the spectral Jacobian together. In particular, an unchanged operator $z_\epsilon=1$ would give power $\epsilon^{a-3}$ in the unnormalized moment. With $z_\epsilon=\epsilon$ the power is $\epsilon^{a-1}$, as in Eq.~\eqref{eq:probeDoubleScaledHat}. The normalization used to state a finite Mellin observable must therefore be specified along with the large-rank scaling.

The higher-frequency terms can be kept directly through the numerical response. For the massless four-dimensional channels, the full finite-window expression is
\begin{equation}
 \widehat{\mathcal G}^{\mathrm{occ}}_{a,L}(\widehat\Lambda)
 =4\pi\mathcal N_{\varphi C}r_h^2
 \int_0^{\widehat\Lambda}\dd\Omega\,
 \frac{\Omega^{a-1}\mathcal R_L(x,\Omega)}
      {e^{\widehat\beta_C\Omega}-1}.
 \label{eq:fullThermalMoment}
\end{equation}
This quantity equals $\epsilon^{1-a}\mathcal G^{\mathrm{occ}}_{\Delta,\Delta';LM}$ within the classical probe family, with the same finite frequency window as Eq.~\eqref{eq:probeMellinBeforeScaling}. It retains the angular transmission and its frequency dependence. The contraction makes this coefficient finite for every fixed window; it leaves the dimensionless product $\widehat\beta_C\widehat\Lambda$ unchanged. The low-frequency approximation consequently has a separate physical scale requirement, which can be assessed from Fig.~\ref{fig:scalarResponse}.

For real $a>1$, the familiar infinite Bose moment is $M(a;\infty)=\Gamma(a)\zeta(a)$. The finite-window tail obeys
\begin{equation}
 0\leq\Gamma(a)\zeta(a)-M(a;X)
 \leq\frac{\Gamma(a,X)}{1-e^{-X}},
 \label{eq:BoseMomentTail}
\end{equation}
where $\Gamma(a,X)$ is the upper incomplete gamma function. The infinite moment describes a spectral model whose linear form extends through the thermally occupied interval. In the Schwarzschild-AdS data at $r_h=\ell$, the interval with less than two per cent departure from the linear slope has $\widehat\Lambda\ell=0.2$ and $\widehat\beta_C\widehat\Lambda=0.2\pi$. We therefore retain the incomplete moment for that interval and use the full radial response for wider thermal windows. Eq.~\eqref{eq:fullThermalMoment} retains the measured spectral shape when the window includes finite-frequency structure. In particular, the homogeneous horizon-area coefficient alone fixes the leading small-window moment, while the larger-window moment resolves the frequency dependence of the absorbing geometry.

The same reasoning applies to a finite-clock channel with $\rho_\tau(\Omega)\sim\epsilon^{-\chi}\Omega^\nu R(\Omega)$. For an operator map $z_\epsilon=\epsilon^q$, a Mellin pair of total weight $a$ scales as $\epsilon^{a+2q-2-\chi}$ in a window $\omega=\epsilon\Omega$. The low-frequency power $\nu$ determines the convergence condition and the temperature moment, while the contraction exponent follows from the clock and normalization map. This separates operator normalization from infrared dynamics. For the probe considered here, $q=\chi=\nu=1$, reproducing the power in Eq.~\eqref{eq:probeDoubleScaledHat}.

\subsection{Time translations in the Mellin representation}
\label{subsec:time-shift-soft-OPE}

The insertion action of a time shift is $\delta_f\Phi_s=f(x)\partial_s\Phi_s$. After regulating the Fourier integral so that its endpoint terms vanish, Eq.~\eqref{eq:celestialoperator} gives
\begin{equation}
 \delta_f\Phi_s^\eta(\omega,x)=-i\eta\omega f(x)\Phi_s^\eta(\omega,x),
 \qquad
 \delta_f\mathcal O_\Delta^\eta(x)
 =-i\eta f(x)\mathcal O_{\Delta+1}^\eta(x).
 \label{eq:supertranslationCelestialShiftEta}
\end{equation}
This algebraic weight shift is the Fourier--Mellin representation of a time derivative. Applying it to the sourced identity in Eq.~\eqref{eq:corrward} gives
\begin{equation}
\begin{aligned}
 0={}&-i\sum_i\eta_i f(x_i)
 \left\langle\mathcal O^{\eta_i}_{\Delta_i+1}(x_i)
       \prod_{j\ne i}\mathcal O^{\eta_j}_{\Delta_j}(x_j)\right\rangle\\
 &+\left(\mathscr D_f^{\mathrm{src}}+\mathscr D_f^\rho\right)
 \left\langle\prod_i\mathcal O^{\eta_i}_{\Delta_i}(x_i)\right\rangle.
 \end{aligned}
 \label{eq:localWardCelestial}
\end{equation}
For constant $f$ in the stationary state, the frequency delta function enforces the cancellation of the insertion terms. For angular $f$, the source variation continues to enter. The constant component is generated by the Brown--York energy, whose equilibrium variation is Eq.~\eqref{eq:firstlawward}. The representation therefore carries the same energy normalization as the thermodynamic identity.

In asymptotically flat scattering, the time-shift charge and its radiative completion lead to the soft-graviton Ward relation \cite{Weinberg:1965nx,Strominger:2013jfa,He:2014laa,Strominger:2017zoo}. The weight shift in Eq.~\eqref{eq:supertranslationCelestialShiftEta} supplies the shared algebraic action. An angular current operator product additionally requires the corresponding current definition, angular transformation law and local operator algebra. Eq.~\eqref{eq:localWardCelestial} gives the sourced thermal identity with the data established here. It can be used when those angular structures are supplied by a specific boundary theory.

\subsection{Stationary three-point data}
\label{subsec:threepoint-Mellin-OPE}

A stationary three-point function has a related transform with two independent frequencies. For a channel with two positive-frequency insertions and one negative-frequency insertion, write
\begin{equation}
 \begin{aligned}
 \langle\Phi_1^+(\omega_1,x_1)\Phi_2^+(\omega_2,x_2)
                   \Phi_3^-(\omega_3,x_3)\rangle
 &=2\pi\delta(\omega_3-\omega_1-\omega_2)\,
 \mathcal C_{123}(x_i)\\[-2pt]
 &\quad\times\omega_1^{\nu_1}\omega_2^{\nu_2}\omega_3^{\nu_3}
 F_{123}(\beta_s\omega_1,\beta_s\omega_2).
 \end{aligned}
 \label{eq:threePointSpectralKernel}
\end{equation}
For this spectral form, the Mellin transform is
\begin{equation}
 \begin{aligned}
 \langle\mathcal O_1^+\mathcal O_2^+\mathcal O_3^-\rangle
 &=2\pi\mathcal C_{123}(x_i)\beta_s^{-A}\mathcal M_{123},\\
 \mathcal M_{123}
 &=\int_0^\infty\dd y_1\dd y_2\,
 y_1^{\Delta_1+\nu_1-1}y_2^{\Delta_2+\nu_2-1}
 (y_1+y_2)^{\Delta_3+\nu_3-1}F_{123}(y_1,y_2),\\
 A&=\Delta_1+\Delta_2+\Delta_3+\nu_1+\nu_2+\nu_3-1.
 \end{aligned}
 \label{eq:threePointCelestialResult}
\end{equation}
Finite windows restrict the integration domain according to all three frequency cutoffs. The common clock Jacobian follows independently: if $\Phi_{s,i}=z_i\Phi_{\tau,i}(\epsilon s)$, the stripped three-point spectral kernel acquires the factor $z_1z_2z_3/\epsilon^2$. Mellin integration then gives the power $z_1z_2z_3\epsilon^{\Delta_1+\Delta_2+\Delta_3-3}$ multiplying the corresponding finite-clock moment. The delta function accounts for one inverse-clock factor and the two independent frequency integrals account for the remaining measure. This provides the normalization needed for an interacting probe calculation.

The transformed three-point function is thermal state data. In a local operator expansion its dependence on temperature combines operator-product coefficients, thermal expectation values and the chosen angular structures \cite{Iliesiu:2018thermal}. Extracting an individual local coefficient requires a specified operator basis and the corresponding decomposition. Eq.~\eqref{eq:threePointCelestialResult} gives the transform of a supplied spectral channel with its clock and operator factors fixed. Bulk interaction vertices can determine that channel, while the two-point calculation in this paper already fixes the source convention, absorption factors and KMS weights that such an extension must use. The transformed channel can thus be compared across the contraction sequence while its local operator decomposition remains specified by the boundary theory.

\section{Conclusion and outlook}
\label{sec:conclusion}

We have connected the finite static Hamiltonian contraction of Ref.~\cite{Xu:2026vpj} to a boundary large-rank ensemble, a sourced stress response and a normalized thermal scalar observable. The boundary clock and Newton coupling scale together, giving a finite Brown--York energy while the entropy and inverse generator temperature grow. The type IIB and BTZ examples fix the relation between the gravitational normalization and microscopic color or central-charge data. The use of independent boundary volume and normalization variables then identifies the work induced by a change of the AdS scale.

The Brown--York result reproduces the finite extended thermodynamic differential in the common-source subtraction scheme. Its volume and normalization contributions combine into the cosmological-pressure work. The finite boundary free energy is the product of the holographic normalization and its conjugate chemical potential, so the established Hawking--Page crossing has a finite boundary thermodynamic interpretation. This relation follows from the weighted homogeneity of the equation of state and the specified source radius. It also identifies the variables that must be differentiated when comparing neighboring boundary ensembles \cite{Karch:2015rpa,Mancilla:2024spp,Borsboom:2026ash}.

The source variation determines the local conservation equations. The spatial terms contain the contraction parameter, and the normalized energy flux has a different weight from the covariant momentum density. At fixed selected time the leading response is stationary, while finite-frequency transport remains visible on the thermal time scale. The global energy is the constant mode of the smeared Brown--York charge. Its conservation and equilibrium variation refer to time evolution and solution-space differentiation, respectively. Angular time shifts transform the source data, and their thermal Ward identity retains those variations.

For the scalar probe, preserving the linear source coupling fixes the operator map between the finite Lorentzian and selected clocks. Together with the Fourier Jacobian, this determines the spectral normalization and the finite Mellin power. Horizon flux supplies the absorptive coefficient, and numerical radial solutions resolve its angular and frequency dependence. At the Hawking--Page radius the zero-frequency dipole and quadrupole transmissions are $0.4549$ and $0.1180$. They approach the homogeneous response as the horizon grows, with the leading correction determined by Eq.~\eqref{eq:largeRadiusTransmission}. The finite-frequency calculation quantifies where the horizon-area slope describes the response and where the full spectral function contributes to the thermal moment.

The type IIB four-point calculation probes the longer logarithmic time interval on which the growing horizon boost compensates the rank suppression. Within the curved-horizon Regge description, the Einstein limit requires the 't Hooft coupling to grow faster than the square of the rank logarithm, while remaining below the string-loop scale. A marginal family retains finite transverse spreading as the string length vanishes. We obtain its spatial kernel, its horizon-wave-function convolution and the relation between integrated attenuation and mean square radius. The instantaneous Regge growth rate approaches the Einstein value throughout this family, whereas the accumulated four-point interaction retains a continuous dependence on the coupling trajectory. The clock and color normalization therefore select observable limits beyond those fixed by the equilibrium equation of state.

The finite-window Fourier--Mellin observable retains that spectral information and obeys the KMS normalization of the contracted ensemble. The incomplete Bose moment is its low-frequency form, and the full radial response determines the same observable through the larger thermal interval. Time translations shift the Mellin weight, with the angular source terms carried into the transformed Ward relation. The stationary three-point formula fixes the corresponding frequency and operator factors for an interacting channel. Its interpretation as thermal data remains tied to the state and to any separately supplied local operator decomposition.

Cubic bulk interactions provide a direct continuation of the stationary response calculation. Evaluating them with the source normalization fixed here would determine the three-point spectral kernels and their dependence on the black-hole state. Time-dependent boundary sources would likewise connect the normalized stress response to transport on the stretched thermal interval. The present construction supplies the clock map, thermodynamic differential and two-point absorption data needed for these calculations. Together they provide a boundary realization of the static Carrollian thermodynamic sector with thermal two-point data and an interacting four-point limit.

\appendix
\section{Boundary susceptibility matrix and contraction conventions}
\label{app:conventions}

The boundary differential can be checked through its mixed derivatives on the enlarged space $(S_C,\mathcal V_\partial,\NC)$. This also identifies the fluctuation directions at fixed spatial volume. Let $n=d-1$, so $S_C=(\Omega_n/4)\NC x^n$ and $\ell=(\mathcal V_\partial/\Omega_n)^{1/n}$. We regard $S_C>0$, $\mathcal V_\partial>0$ and $\NC>0$ as independent coordinates when taking the following derivatives. Restricting to fixed $G_C$ is performed after this calculation. The contraction parameter remains fixed during all thermodynamic differentiation.

At fixed $\mathcal V_\partial$, simultaneous scaling of $S_C$ and $\NC$ leaves $x$ unchanged and scales the energy linearly. The temperature and color chemical potential therefore depend on their ratio. Differentiating the equation of state gives
\begin{equation}
\begin{aligned}
 \left(\frac{\partial\widehat T_C}{\partial\NC}\right)_{S_C,\mathcal V_\partial}
 &=\left(\frac{\partial\mu_{\NC}}{\partial S_C}\right)_{\NC,\mathcal V_\partial}\\
 &=\frac{(d-2)-d x^2}{4\pi\ell(d-1)\NC x}.
 \end{aligned}
 \label{eq:mixedTemperatureColor}
\end{equation}
This relation checks the color work independently of the fixed-$G_C$ bulk curve. Increasing the holographic normalization at fixed entropy decreases $x$. On the large-black-hole branch it lowers the normalized temperature, with the sign change occurring at the branch merger. The derivative consequently measures how the same total entropy is distributed across a varying number of degrees of freedom.

The Hessian restricted to entropy and holographic normalization has a simple form:
\begin{equation}
 \left.\operatorname{Hess}E_C\right|_{\mathcal V_\partial}
 =\frac{\widehat T_C}{C_C}
 \begin{pmatrix}
 1 & -S_C/\NC\\
 -S_C/\NC & S_C^2/\NC^2
 \end{pmatrix}.
 \label{eq:entropyColorHessian}
\end{equation}
Its null direction is proportional to $(S_C,\NC)$, corresponding to the common normalization change at fixed $x$. The other direction has the sign of $C_C$, since $\widehat T_C>0$. The locally stable black-hole branch therefore has positive curvature in the direction that changes entropy per degree of freedom. The small-black-hole branch has negative curvature in that direction. The color derivative and heat capacity are thus linked by the same equation of state, with the zero direction fixed by normalization homogeneity.

Volume derivatives provide another check. The boundary pressure satisfies $p_\partial=E_C/(n\mathcal V_\partial)$, while both $\widehat T_C$ and $\mu_{\NC}$ scale as $\mathcal V_\partial^{-1/n}$ at fixed $x$. Consequently,
\begin{equation}
\begin{aligned}
 \left(\frac{\partial\widehat T_C}{\partial\mathcal V_\partial}\right)_{S_C,\NC}
 &=-\left(\frac{\partial p_\partial}{\partial S_C}\right)_{\mathcal V_\partial,\NC},\\[4pt]
 \left(\frac{\partial\mu_{\NC}}{\partial\mathcal V_\partial}\right)_{S_C,\NC}
 &=-\left(\frac{\partial p_\partial}{\partial\NC}\right)_{S_C,\mathcal V_\partial}.
 \end{aligned}
 \label{eq:volumeMaxwellRelations}
\end{equation}
These identities ensure that the source-volume and color terms belong to one energy potential. The restriction $\delta\log\mathcal V_\partial=\delta\log\NC$ at fixed $G_C$ then reproduces the bulk pressure coefficient. A different vacuum-energy convention adds its derivatives to all of these response functions; the common AdS subtraction used in the main text fixes that convention throughout.

For the real-time observables, the relevant variable change is $s$ is the selected-generator clock, $\tau=\epsilon s$ is the finite Lorentzian clock, and $\Omega=\omega/\epsilon$ is held fixed when the frequency window contracts. The source-normalized scalar in Eq.~\eqref{eq:scalarSourceMap} is the response density with respect to $\dd s\sqrt h$. A scalar operator defined through the full covariant measure $\sqrt{-g}\,\dd s\,\dd^{d-1}x$ instead assigns the lapse factor to the measure. Stating the measure and source map together determines which operator carries the factor of $\epsilon$. Our spectral and Mellin formulae use the former density convention, with the same $\epsilon$ in the Gibbs period and Fourier Jacobian. This fixes the operational meaning of every frequency and operator factor in the reported thermal response.

\bibliographystyle{JHEP}
\bibliography{reference}
\end{document}